\RequirePackage{ifpdf}
\documentclass[11pt,a4paper]{article}
\usepackage{jcappub}
\usepackage{amsmath}
\usepackage{dcolumn}
\usepackage{breqn}
\usepackage{bm}
\usepackage{color}
\usepackage{cancel}
\usepackage[utf8]{inputenc}
\usepackage{graphicx}
\usepackage{caption}
\usepackage{subcaption}
\usepackage{tabularx}

\usepackage{ulem}

\newcommand{\beq}{\begin{equation}}
\newcommand{\eeq}{\end{equation}}

\newcommand{\mpl}{M_{\rm Pl}}
\title{Non-Gaussianities in the Extended EFT of Inflation}

\author[a]{Amjad Ashoorioon,}
\author[b,e]{Ghazal Geshnizjani,}
\author[b]{Hyung J. Kim}

\affiliation[a]{School of Physics, Institute for Research in Fundamental Sciences (IPM),\\
P.O. Box 19395-5531, Tehran, Iran}
\affiliation[b]{Department of Applied Mathematics, University of Waterloo\\ Waterloo, Ontario, N2L 3G1, Canada}
\affiliation[c]{Perimeter Institute for Theoretical Physics\\ 31 Caroline St. N., Waterloo, ON, N2L 2Y5, Canada}

\emailAdd{amjad@ipm.ir}
\emailAdd{ggeshniz@uwaterloo.ca}
\emailAdd{h268kim@uwaterloo.ca}

\abstract{In earlier works, we studied the validity of Extended Effective Field Theory of Inflation (EEFToI) in the regime where initial conditions are set with dispersion relations $\omega^2 \propto k^6$. We had also evaluated and examined the power spectrum for some interesting corners of the parameter space. In this paper, we compute the bispectrum in the EEFToI, take a closer look at the strong coupling constraints and calculate the size of the non-Gaussianities in those regions of parameter space. We also investigate the shape of triangles that contribute to the enhancement of non-Gaussianities in this regime. We find that there are allowed parts of parameter spaces where EEFToI description with initial conditions set with $\omega^2 \propto k^6$ is sensible and interesting.}


\subheader{IPM/P-2021/33}

\begin{document}
\maketitle

\section{Introduction}

The inflationary cosmology is one of the most widely accepted paradigms of the early universe cosmology. An early phase of quasi-de Sitter expansion in which the universe accelerated, was  developed in order to address some existing issues of the Standard Big Bang (SBB) Cosmology, such as the flatness and the horizon problem\cite{Linde:1986fc, Guth:2007ng}. Soon it was also recognised that it can also provide a framework for generating the seeds of large scale structures (LSS) from sub-horizon vacuum quantum fluctuations \cite{Mukhanov:1981xt}. However, with all its successes it is important to note Inflation is not just one specific model manifested from a fundamental theory of particle physics and gravity. In fact, as we do not yet have a well understood theory of quantum gravity, every inflationary model should be viewed as an Effective Field Theory in the semi-classical gravity regime. Therefore, it is important to make sure the effective field theory description is self consistent in its regime of its validity \cite{Babic:2019ify, Burgess:2017ytm}. Meanwhile, as a healthy scientific exercise we should also explore if there are other sensible alternative scenarios \cite{Kim:2020iwq,Boruah:2018pvq} to inflationary paradigm that can describe the initial conditions of the universe. This is specially relevant in the light of some of the less desirable outcomes of the inflation such as eternally inflationary spacetimes\cite{Linde:1986fc, Guth:2007ng} or being geodesically past incomplete \cite{Borde:1993xh}. The latter implies inflation does not address the singularity problem \cite{Borde:2001nh}.
\par As for probing Inflation through a consistent effective field theory framework, which is the topic of this paper, in 2006 Creminelli et al. \cite{Creminelli:2006xe} introduced a systematic way to write the effective field theory for the cosmological perturbations of single field models without explicitly specifying the details of the action of the matter field. In this formalism, coined as the Effective Field Theory of Inflation (EFToI), the action for curvature perturbations is constructed by requiring the original theory to be four dimensional diffeomorphism invariant with only one dynamical scalar degree of freedom (inflaton field). The action is constructed in the unitary gauge, {\it i.e.} the preferred spatial slicing of the inflaton, where it is taken to be the clock. In this gauge, the time diffeomorphism is spontaneously broken and is no longer a linearly realized symmetry. Due to the reduced symmetry in this gauge, there can be more terms in the action as all the 3D spatial diffeomorphism invariant terms can also be included. This action is further simplified by expanding it perturbatively around a quasi-de Sitter Friedman-Robertson-Walker (FRW) background and in the decoupling limit, where mixing of the scalar mode with gravity can be ignored. The detailed derivation of the action will be reviewed later in the next section. One of the strengths of this approach is that it can unite and reproduce all the other single field inflationary models that had previously been encountered such as the canonical single field model, ghost inflation, and DBI inflation, by adjusting the parameters under one framework. In particular, it allows exploring some of the non-standard features of single field models, such as the deviations of the speed of propagation from speed of light or modified dispersion relations in light of the newly introduced operators in unitary gauge. 
\par Having noticed the potential of describing higher order dispersion relations within EFToI, the very early works of EFToI \cite{Cheung:2007st,Creminelli:2006xe,Bartolo_2010}, explored their implications for inflation up to the order of $\omega^2 \propto k^4$. Nevertheless, terms leading to $\omega^2 \propto k^n$ with $n\geq6$ were not studied, due to the speculations of strong coupling at low energy limit. However, as pointed out in our papers \cite{Ashoorioon:2017toq,Ashoorioon:2018uey,Ashoorioon:2018ocr} using a power counting argument for all the the relevant operators, EFT does not necessarily face any strong coupling problem in the low energy as it naturally renders back to the $k^4$ and $k^2$ regimes, when the mode gets stretched due to the universe expansion. Therefore, we introduced the notion of Extended Effective Field Theory of Inflation (EEFToI) where operators that could produce sixth order $k$ terms or higher orders can also be included in the action. Our focus in those works was primarily on the implications of such models for the two point function and we left out the rigorous calculation of the three point function for later. In particular, we examined the parameter space of the modified dispersion relations of EEFToI, where adiabatic conditions were mildly violated but still under control, and computed the enhancement to the amplitude of the curvature perturbations compared to the cases initiating from the Bunch Davies initial conditions. This analysis allowed us to find some regions in the parameter space that can lead to one to two orders of magnitude enhancement of the scalar power spectrum which in return imply lower energy scales for inflation itself \footnote{This framework can also be further pushed into the regions of the parameter space where the power spectrum gets modulated by many more orders of magnitude and $\Delta{{}_{\zeta}}\sim 10^{-1}$ at the relevant scales, to explain the origin of enhancement in the scalar power spectrum needed to explain the formation of primordial black holes (PBHs) \cite{Ashoorioon:2019xqc}. There it is argued that the effective field theory criterion of avoidance of the strong coupling could be satisfied too, as the modes for such dispersion relations are at most dominated by the quartic part of the dispersion relation at horizon crossing. However, in this work, we mainly focused on the regions of the parameter space that was discussed in \cite{Ashoorioon:2018uey} where the enhancement is at most about an order of magnitude from the usual Lorentzian dispersion relation with the Bunch-Davies vacuum.}. Having done that study, we now return to the analysis of the three point function and the strong coupling constraints, to obtain a more clear picture of the tuning required for the coefficients of the higher dimensional operators in order to have a sensible EFT description with initial conditions set via $\omega^2 \propto k^6$ at early times. In fact since the observation of the fluctuations in the cosmic microwave background (CMB) and the measurements of $f_{{}_{\rm NL}}$ parameters impose a much tighter constraints on interaction terms in the action, we use them to further scrutinise the effects of the modifications to dispersion relations \footnote{For study of non-Gaussianity in the EFToI refer to \cite{Bartolo_2010,Cheung:2007st}}. Furthermore, if the primordial non-Gaussianities are eventually observed, calculating the non-Gaussianity signatures of the theory allows us to distinguish between different scenarios of inflation. 

The outline of the paper is as follows. In section \eqref{section:EFT_Review}, we will briefly review EFToI and its extension to EEFToI. In section \eqref{section:ng}, we will examine the interaction terms and calculate the bispectrum. Section \eqref{sec:eeftoiresult} summarises our numerical result for non-Gausianities and constraints on parameter space. We will make our concluding remarks in section \eqref{sec:conc}. 

\section{Review of Effective Field Theory and Extended Effective Field Theory of Inflation}\label{section:EFT_Review}
In the context of inflationary cosmology, the basic idea of the single field EFToI is that even though the action is invariant under all diffeomorphisms (temporal and spatial), the cosmological solution has a preferred time slicing, in which the fluctuations of the inflaton field vanish. In this gauge, which is known as the unitary gauge, the surviving symmetries of the action are only the spatial diffeomorphisms. In the unitary gauge, the dynamical scalar degree of freedom which was sourced by the scalar field, is manifested in the longitudinal component of the metric through the Goldstone boson corresponding to the broken time symmetry. As expected this Goldstone boson transforms non-linearly under the time diffeomorphism and one can restore the time diffeomorphism back in the action, using the Stueckelberg trick. This property provides a framework to write down the most general theory of fluctuations around a time-evolving background where the time diffs are non-linearly realized.

To start, one writes the most general action in unitary gauge using all the possible operators that satisfy spatial diffeomorphisms. These operators, in addition to the 4D diff invariant terms such as Reimann tensor and its covariant derivatives, can also include terms that are pure functions of time, extrinsic curvature terms $K_{\mu\nu}$ and $g^{00}$ operators. It has been shown in details that using this formalism, the Lagrangian of single scalar field inflationary models around flat Friedmann-Lema\^itre-Robertson-Walker (FLRW) metric can be written as \cite{Cheung:2007st,Ashoorioon:2018uey,Ashoorioon:2018ocr,Ashoorioon:2018ocr}
\begin{equation}
\mathcal{L}
=
M_{{{}_\rm Pl}}^2
\left[
\frac12\, R
+\dot H\,  g^{00}
-\left(3\, H^2 +\dot H\right)
\right]
+\sum_{m\geq 2}\,
\mathcal{L}_m (g^{00}+1,\delta K_{\mu\nu}, \delta R_{\mu\nu\rho\sigma};\nabla_\mu;t)
\ ,
\label{genspacei}
\end{equation}
where $\mathcal{L}_m$ represents the spatially invariant terms of perturbative order $m$ in arguments $g^{00}+1,~\delta K_{\mu\nu}$ and $\delta R_{\mu\nu\rho\sigma}$. This action can in principle include many terms each describing perturbations in a different inflationary model. However, in the original derivation of EFToI \cite{Cheung:2007st}, due to the reasons that will be explained soon, only the terms of mass dimensions up to 2 were considered in the sum. Later, in the EEFToI \cite{Ashoorioon:2018uey,Ashoorioon:2018ocr}, the terms up to $[{\rm mass}]^4$ were also examined and one could even argue more terms can be included too. Once a selection of terms are written down, then the St\"{u}ckelberg trick is used to make the Goldstone boson manifest. For that, first the time transformation $t \rightarrow t+\xi^0({x^\mu})$ is taken and then time diffeomorphisms was restored by substituting $\xi^0(x^\mu)$ with a field $-\pi(x^\mu)$. Now asserting that under time diff $\pi \rightarrow \pi-\xi^0({x^\mu})$, action becomes invariant under all diffs at all orders. Using this procedure, one can retrieve the Lagrangian density of the canonical single field slow-roll inflation from the EFToI by taking the terms of perturbative order of $m=0$. That entails, expanding $m=0$ terms, perturbatively in $\pi$ and taking the decoupling limit where $\dot{H}\rightarrow0$ while $M_{{}_{\rm Pl}}^2\dot{H}$ is kept fixed, leading to the following Lagrangian density for $\pi$ \footnote{where we have chosen $(-,+,+,+)$ convention}, 
\begin{equation}
    \mathcal{L}_{{}_{\rm slow-roll}}= -M_{{}_{\rm Pl}}^2\,  \dot{H}\left(\dot{\pi}^2-{\frac{(\partial_i \pi)^2}{ a^2}}\right)\,.
\end{equation}
One can substitute $\pi$ with the more familiar conserved quantity $\zeta = -H\pi$ to obtain the standard slow-roll inflationary action for $\zeta$.
\par Next, including more terms with $m\geq 2$ and operators up to mass order two ($d=2$) a larger class of inflationary scenarios were identified and studied \cite{Cheung:2007st},
\begin{eqnarray}
\label{eftoiunigauge}
\mathcal{L}_{{}_{\rm EFToI}}&=&M_{{}_{\rm Pl}}^2
\left[
\frac12\, R
+\dot H\,  g^{00}
-\left(3\, H^2 +\dot H\right)
\right]\nonumber\\
&&+
\frac{M_2^{4}}{2!}\,(g^{00}+1)^2+
\frac{M_3^{4}}{3!}\,(g^{00}+1)^3
+\frac{\bar M_1^{3}}{2}(g^{00}+1) \delta K^\mu_{\ \mu} -\frac{\bar M_2^{2}}{2}\, (\delta K^\mu_{\ \mu})^2
\nonumber\\&&
-\frac{\bar M_3^{2}}{2}\, \delta K^\mu_{\ \nu}\,\delta K^\nu_{\ \mu}
\ .
\end{eqnarray}

Here, the upper indices are exponents of $M^{i}$ and $\bar{M}^{i}$ and are adjusted to reflect the mass dimension of the coefficients. 
Note that the Lagrangian density above is not an exhaustive list of all the possible operators with mass dimensions $d\leq 2$. Since $(g^{00}+1)$ is dimension zero, multiples of the terms listed above and powers of $(g^{00}+1)$ will not change their mass dimension. Since $g^{00}+1\sim 2\dot{\pi}$, these new terms will not contribute to powerspectrum but may change bispectrum or higher order correlation functions. There is also another mass dimension squared operator, $\bar{M}^2\nabla^{\mu}g^{00}\nabla_{\mu}g^{00}$ which after the implementation of the Stueckelberg technique yields Ostragradski ghosts at second order for the Goldstone boson, $\pi$. That operator in the unitary gauge action leads to
\begin{equation} \label{EFToIaction}
    \bar{M}^2 \left(-\ddot{\pi}^2+\frac{k^2}{a^2}\dot{\pi}^2\right)\,.
\end{equation}
in the Lagrangian for the Goldstone  boson. As we will see, such ghosts are generated in the Extended EFT of inflation too which limits the size and value of their corresponding coupling constants \footnote{The second term, $\frac{k^2}{a^2}\dot{\pi}^2$ leads to time-dependent modification of the speed of sound, making it vanish at infinite past, when the wavelength of the modes are small. This will lead to reducing the cutoff below the Hubble parameter at some point in the evolution of the modes and lead to strong coupling, please see \cite{Ashoorioon:2018ocr} for a similar situation that arises in the context of Extended EFT of inflation.}.

The Lagrangian of the EFToI \eqref{eftoiunigauge} in the decoupling limit, where $\dot{H}\rightarrow0$ and $M_{{}_{\rm Pl}}^2\dot{H}$ is kept fixed, results into the following quadratic and cubic expressions for $\pi$ \cite{Cheung:2007st, Bartolo:2010bj}:
\begin{eqnarray}
\mathcal{L}_{{}_{\rm EFToI}}^{(\pi)}&=& -M_{{}_{\rm Pl}}^2\,  \dot{H}\left(\dot{\pi}^2-{\frac{(\partial_i \pi)^2}{a^2}}\right)+2 M_2^4 \left( \dot{\pi}^2+\dot{\pi}^3
-\dot{\pi}\frac{(\partial_i \pi)^2}{2 a^2}\right)-\frac 43M_3^4 \dot{\pi}^3 -\bar M_1^3\bigg(3H\dot{\pi}^2\nonumber\\
&-&H\frac{(\partial_i \pi)^2}{2 a^2}
-\frac{(\partial_j^2\pi)(\partial_i \pi)^2}{2 a^4}\bigg)-\frac{\bar M_2^2}{2} \bigg(9H^2 \dot{\pi}^2
-3 H^2 \frac{(\partial_i \pi)^2}{a^2}+\frac{(\partial_i^2 \pi)^2}{a^4} + \frac{H(\partial_i^2\pi)(\partial_j\pi)^2}{a^4}\nonumber\\
&+&\frac{2\dot\pi\partial_i^2\partial_j\pi\partial_j\pi}{a^4}\bigg)-\frac{\bar M_3^2}{2} \bigg(3H^2 \dot{\pi}^2- H^2 \frac{(\partial_i \pi)^2}{a^2}+\frac{(\partial_i^2 \pi)^2}{a^4} + \frac{2H(\partial_i^2\pi)(\partial_j\pi)^2}{a^4}
\nonumber\\&+&\frac{2\dot\pi\partial_i^2\partial_j\pi\partial_j\pi}{a^4}\bigg)
+\dots\ . \nonumber\\
\label{Lpi}
\end{eqnarray} 
First, note that any term that contributes to the coefficient of $\dot{\pi}^2$ in the approximate plain wave regime, leads to deviations from $c_{{}_S}=1$. However, any terms that results in spatial derivatives for $\pi$, can potentially modify the dispersion relation. Some modifications could even lead to non-hyperbolic equations of motion for $\pi$ that do not have propagating solutions or are not even well posed. As we see, the first term in above expression, $\frac{M_2^4}{2!}\,(g^{00}+1)^2$, only contributes to modifying $c_{{}_S}$ \cite{Cheung:2007st},
\begin{equation}\label{cssqyared}
c_{{}_S}^{-2}=1-\frac{2M_2^4}{\mpl^2 \dot{H}}\,.
\end{equation}
This shows that including only this term,  the superluminal propagation can be avoided for $M_2^4>0$, since during inflation $\dot{H}<0$  \footnote{There are some arguments that superluminal propagation could lead to the prohibition of UV completion \cite{Adams:2006sv}.}. Through a quick substitution for $M_2$, we see the cubic term in $\pi$ from this same operator, appears with a factor proportional to $(1-\frac{1}{c_{{}_S}^2})$. This agrees with the generic feature that in the single field inflationary models, smaller values of the speed of sound result in the enhancement of the non-gaussianities. This in return, through observational bounds, further constraints the value of $M_2$. 

Next, the term proportional to $\bar{M}_1$ in \eqref{eftoiunigauge} is odd under time reversal and if we assume theory has time reversal symmetry we could set it to zero. Nevertheless, it also preserves the form of dispersion relation as $\omega^2=c_{{}_S}^2 k^2$ but contributes to changes in speed of sound. In the exact de-Sitter limit, $\dot{H}=0$, with $M_2\neq0$ and $\bar{M}_1\sim - M_2\sim \bar{M}$, the speed of sound turns out to be quite small, $c_{{}_S}^2=\frac{H}{4 \bar{M}}\ll 1$, but if $M_2=0,~ M_3<0$, then $c_{{}_S}^2=\frac{1}{6}$.

Finally, the last two terms would not only add corrections to the sound speed for $\pi$ field, but also induce corrections proportional to $k^4$ to the dispersion relation. The effect of such quartic corrections to the dispersion relation on the power spectrum and bispectrum has been worked out both analytically and numerically \cite{Ashoorioon:2011eg,Bartolo:2010bj}.

While in the original work of \cite{Cheung:2007st}, it was already noticed that including other possible operators in \eqref{genspacei} in the action, could lead to dispersion relations beyond the quartic form, it was argued that those possibilities would lead to the violation of the EFT criterion. The rationale of the argument is that, assuming $\omega^2\propto k^{2n}$ with $n\geq3$, the power of the energy scaling of $\pi$ would be $-1/2+3/2n$ which means the scaling dimension of the cubic operator $\dot{\pi} (\partial_i\pi)^2$ would be $(7-3n)/(2n)$. For $n\geq3$, {\it i.e.} beyond quartic form, this scaling dimension becomes negative which leads to this operator becoming relevant at low energies and hence the effective field theory regime breaks down. However, as we argued in \cite{Ashoorioon:2018uey} this statement would be true if $\omega^2\propto k^{2n}$ dispersion relation remained valid up to horizon crossing and there were no other intermediate EFT regime that dominated the dispersion relation at low energies. We pointed out first, interestingly there are always finite number of relevant operators for every given $n$. Second, the operators $(\partial_i^2 \pi)^2$ and $(\partial_i \pi)^2$ that change the dispersion relation are among relevant operators with negative scaling but larger in magnitude powers. Therefore, they can make the theory transition into a different regime of dispersion relation before it becomes strongly coupled. 
For example, in the case of $\omega^2\sim k^6$, the scaling power of $\dot{\pi} (\partial_i\pi)^2$ is $-1/3$, but the scaling power of $(\partial_i^2\pi)^2$ and $(\partial_i\pi)^2$ are respectively $-2/3$ and $-4/3$. This implies that the dispersion relation would change from $\omega^2\sim k^6$ to $\omega^2\sim k^4$ or $\omega^2\sim k^2$ at some energy scale $\Lambda_{\rm dis}$. Hence, if the coefficients of the higher dimensional operators are such that different cutoffs of each EFT regime follow a hierarchy of successive scales, theory remains weakly coupled throughout its evolution down to low energies\cite{Ashoorioon:2018uey}. This in principle can allow for an interesting extension of the EFT  regime to $\omega^2\propto k^{6}$.

In the case of EFT of inflation, the extrinsic curvature squared terms in the action not only induced corrections proportional to $k^4$, but they induce time dependent corrections to $k^2$. Therefore, pure $k^4$ dispersion relation can come about only through fine-tuning the coefficients of the operators in the unitary gauge action. Likewise, addition of higher mass dimension operators in the unitary gauge action, not only produces corrections beyond the quartic order to the dispersion relation, they will at the same time induce lower order corrections to the dispersion relation which will become important at the low energy scales. Hence, it seems that the theory benefits from some self-healing properties.

Motivated by this reasoning, in the EEFToI papers \cite{Ashoorioon:2017toq,Ashoorioon:2018uey}, we extended the EFToI formalism to study the consequences of including terms up to order $d=4$ that can further change the dispersion relation in addition to the terms that had previously been studied in EFToI:
\begin{equation}
\mathcal{L}_{{}_{\rm EEFToI}} = \mathcal{L}_{{}_{\rm EFToI}}+\sum_{d=3}^4\mathcal{L}_{2, d} 
\label{LEFT}
\end{equation}
The additional terms up to order $d=4$ that were included in EEFToI are given by
\begin{eqnarray}
\label{eq:actiontad}
\sum_{d=3}^4\mathcal{L}_{2, d}
&=&
\frac{\bar{M}_4}{2} \nabla^\mu g^{00}\nabla^\nu\delta K_{\mu \nu}-\frac{\delta_1}{2}\,(\nabla_{\mu} \delta K^{\nu\gamma})(\nabla^ {\mu} \delta K_{\nu\gamma})
-\frac{\delta_2}{2}\,(\nabla_{\mu} \delta K^\nu_{\ \nu})^2
\nonumber\\
&&-\frac{\delta_3}{2} \,(\nabla_{\mu} \delta K^\mu_{\ \nu})(\nabla_{\gamma} \delta K^{\gamma\nu})-\frac{\delta_4}{2} \,\nabla^ {\mu}\delta K_{\nu\mu}\nabla^ {\nu}\delta K_{\sigma}^{\sigma}.
\end{eqnarray} Here, $\delta_i$ are dimensionless coefficients. 

The new terms included in the Extended EFToI \eqref{eq:actiontad} for mass dimension 3 and 4, lead to the following corrections to the quadratic action in Foureir space: 
\begin{eqnarray}
\mathcal{L}_{2, 3}^{(\pi_k)} = \frac{\bar M_4}{2}\left(\frac{k^4 H \pi_k^2}{a^4}+\frac{k^2 H^3 \pi_k^2}{a^2}-9 H^3\dot\pi_k^2\right)\,,
\label{Ld3}
\end{eqnarray} and
\begin{eqnarray}\label{action-pi}
\mathcal{L}_{2, 4}^{(\pi_k)}&=&-\frac{1}{2} \delta_1 \left(\frac{k^6 \pi_k ^2}{a^6}-\frac{3 H^2 k^4 \pi_k ^2}{a^4}-\frac{k^4 \dot{\pi}_k^2}{a^4}+\frac{4 H^4 k^2 \pi_k ^2}{a^2}-6 H^4 \dot{\pi}_k^2-3 H^2 \ddot{\pi_k}^2\right)\nonumber\\&&
-\frac{1}{2} \delta_2 \left(\frac{k^6 \pi_k ^2}{a^6}+\frac{H^2 k^4 \pi_k ^2}{a^4}-\frac{k^4 \dot{\pi}_k^2}{a^4}+\frac{6 H^4 k^2 \pi_k ^2}{a^2}-9 H^2 \ddot{\pi_k}^2\right)\nonumber\\&&
-\frac{1}{2} \delta_3 \left(\frac{k^6 \pi_k ^2}{a^6}+\frac{3 H^2 k^4 \pi_k ^2}{a^4}+\frac{ H^2 k^2 {\dot{\pi}_k} ^2}{a^2}-9 H^4 \dot{\pi}_k^2\right)\nonumber\\&& -\frac{1}{2}\delta_4 \left(\frac{k^6 \pi_k ^2}{a^6}+\frac{ H^2 k^4 \pi_k ^2}{2 a^4}+\frac{9 H^4 k^2 \pi_k ^2}{2 a^2}+\frac{3 H^2 k^2 \dot{\pi}_k^2}{a^2}+\frac{27}{2} H^4 \dot{\pi}_k^2\right)\,. \nonumber\\
\end{eqnarray}
As we see $\delta_1$ and $\delta_2$ expressions contain $(\ddot\pi)^2$  or Ostrogradsky ghosts. In the context of an effective field theory, ghosts are not necessarily a pathology if they are small corrections and under control. This is because ghosts maybe an artifact of the truncation and the effective theory is only meant to capture the relevant features of the full theory \cite{Burgess:2014lwa, Garriga:2012pk}. However, in this work we will not study their consequences and for simplicity focus on the ghost free possibilities by setting the coefficients $\delta_1 = \delta_2 = 0$ \footnote{Setting $\delta_1+3\delta_2=0$, one could in principle avoid the $\ddot{\pi}^2$ in the action for the Goldstone boson, $\pi$. The term proportional to $\delta_1$, $(\nabla_{\mu} \delta K^{\nu\gamma})(\nabla^ {\mu} \delta K_{\nu\gamma})$, gives rise to ghosts for the tensor perturbations too \cite{Ashoorioon:2018uey}. Imposing lack of ghosts in both the scalar and tensor sectors of the theory one has to impose $\delta_1=\delta_2=0$} \footnote{One can show that if $\nabla^{\mu}$ operators are contracted with the induced metric, $h^{\mu\nu}$, there will be no ghosts in the scalar and tensor part of the theory. However, here we keep to the original framework of the Effective Field Theory of inflation, in which operators built out of $\nabla_{\mu}$ are allowed as they are invariant under spatial diffeomorphisms\cite{Ashoorioon:2018uey}.}.

Also we like to point out that similar to EFToI, \eqref{Ld3} is not the exhaustive list of all $d=3$ and $d=4$ operators. To begin with, we are only including terms that are quadratics in $\delta g^{00}$ and $\delta{K}$, as only those terms lead to corrections for the dispersion relation. Next, in the derivation of the above expression terms that could be absorbed into other ones through integration by part are not included. For example, the terms $(\nabla_\mu\nabla_\nu\delta g^{00}) \delta K^{\mu\nu}$, and $\delta g^{00} (\nabla_\mu\nabla_\nu\delta K^{\mu\nu})$ are absorbed into the $\bar{M}_4$ term. Finally, there are more terms such as $(\nabla_\mu\nabla^\mu\delta g^{00})^2$, $(\nabla_\mu\nabla^\mu\delta g^{00}) \text{Tr}(\delta K)$ that have been discarded as they also result in $(\ddot\pi)^2$ in the quadratic Lagrangian for $\pi$ \footnote{Similarly one can avoid the ghosts in these operators by contracting the $\nabla_{\mu}$ operators with the induced metric, rather then the four dimensional metric.}.  There is also the possibility that in a particular corner of the parameter space, the coefficients of these terms are non-zero, yet they conspire in a way that all the $(\ddot{\pi})^2$ terms cancel out each other and the theory is ghost free. In such a case, assuming there is any interesting physics, contributions to terms such as $k^2\dot{\pi}_k^2$ and $k^4\dot{\pi}_k^2$ and the potential strong coupling regimes need to be studied carefully \cite{Ashoorioon:2018ocr}.

Going forward we also only explore models where $\delta_3 = -3\delta_4$. The reason for this additional assumption is that we are more interested in investigating models where the effects of $k^6 \pi_k ^2$ operators is dominant and setting $\delta_3 = -3\delta_4$ will automatically cancel off $k^2 \dot{\pi}_k^2$ terms. Deviating from this condition, one has to carefully take into account the competing impact of $k^2 \dot{\pi}_k^2$ term which can introduce a different cut off and dispersion relation in UV. However, it can still produce a well-defined finite power spectrum under certain criteria \cite{Ashoorioon:2018ocr}. 

\begin{figure}[ht]
     \centering
     \begin{subfigure}[b]{0.9\textwidth}
         \centering
         \includegraphics[width=\textwidth]{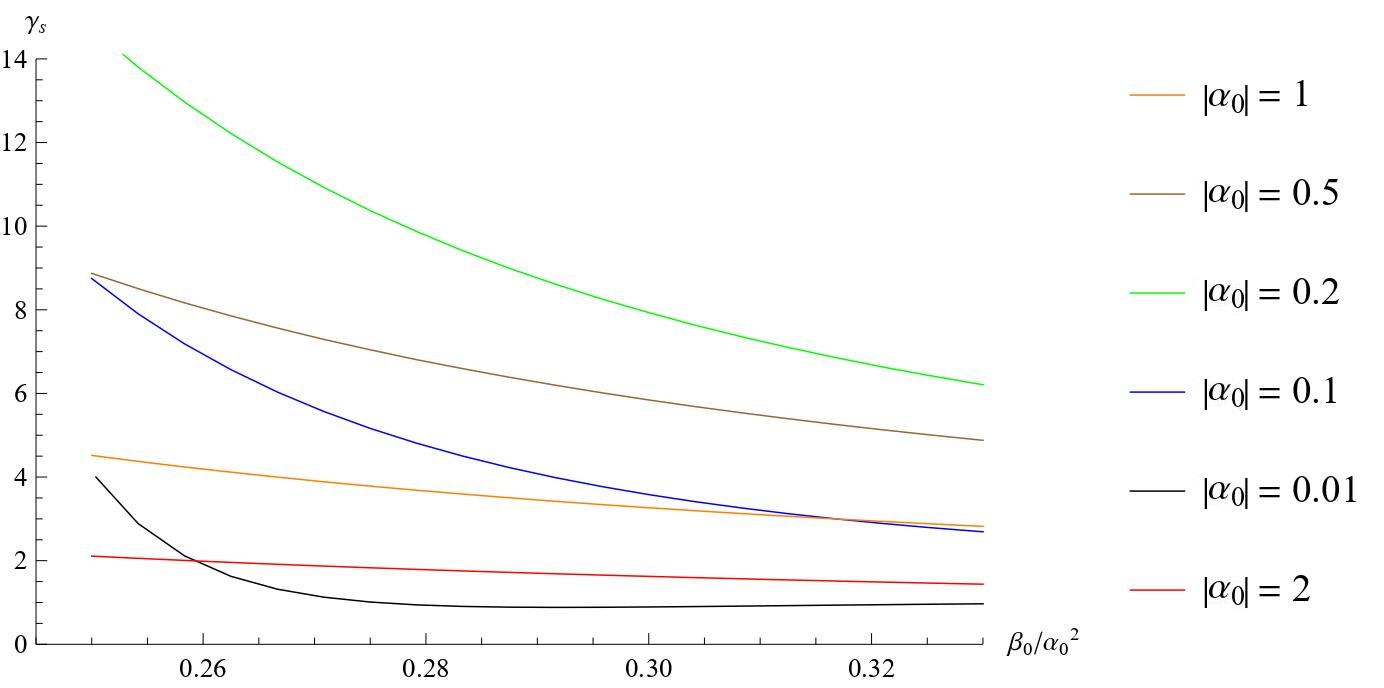}
         \caption{$\gamma_{{}_S}$ as a function of $\beta_0/\alpha_0^2$ when $\alpha_0<0$ and $\beta_0>0$ and $0.01 \leq |\alpha_0| \leq 2$ and $\frac{\alpha^2}{4} \leq \beta_0$.}
     \end{subfigure}

     \begin{subfigure}[b]{0.9\textwidth}
         \centering
         \includegraphics[width=\textwidth]{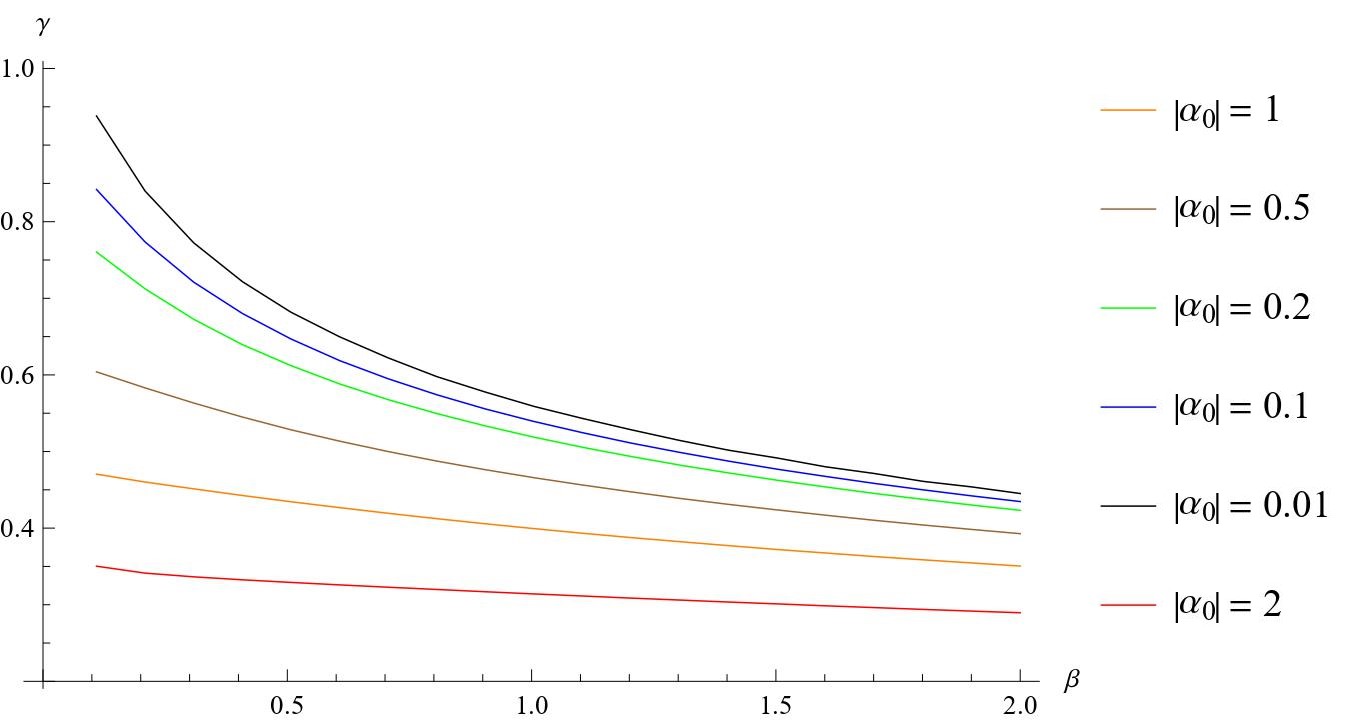}
         \caption{Plot of $\gamma_{{}_S}$  as a function of $\beta_0$ when $\alpha_0$ and $\beta_0$ are both positive and $1\leq \alpha_0\leq2$ with $0\leq \beta_0\leq2$.}
     \end{subfigure}
     \caption{Plot of the amplification of power spectrum compared to the Bunch Davies power spectrum $P_s^{B.S}$ with varying values of $\alpha_0$ and $\beta_0$.} 
    \label{fig:two graphs}
\end{figure}

Under the assumptions mentioned above and after we reorganize all the terms contributing to the Lagrangian, it will have the following form in Fourier space:  
\begin{equation}
\int dt\,d^3k \sqrt{-g}\, \mathcal{L}_{{}_{\rm EEFToI}} =\int dt\, d^3k \, a^3 \left (\frac 12 A_1\dot{\pi}_k^2+ \frac 12 \left [C_1 \frac{k^6}{a^6}+D_1\frac{k^4}{a^4}+F_1\frac{k^2}{a^2} \right ] \pi_k^2\right) ,
\label{KEFT}
\end{equation}
where the the coefficients $A_1, C_1, D_1$ and $F_1$ are defined as 
\begin{eqnarray}\label{coefs}
&&A_1=-2 M_{{}_{\rm Pl}}^2 \dot{H}+4 M_2^4-6\bar M_1^3 H-9 H^2 \bar M_2^2-3 H^2 \bar M_3^2+\frac{27}{2}H^4\delta_3-9H^3 \bar M_4\,,\nonumber\\
&&C_1=\frac 23 \delta_3\,,\nonumber\\
&&D_1=  \bar M_2^2+\bar M_3^2+\frac{17}{18} H^2 \delta_3 -\bar M_4 H\,,\nonumber\\
&& F_1=-2 M_{{}_{\rm Pl}}^2 \dot{H}-\bar M_1^3 H-3H^2 \bar M_2^2-\bar M_3^2 H^2+\frac 12  H^4 \delta_3-\bar{M}_4 H^3
\, .
\end{eqnarray}
Next, expanding the Canonical variable $a\sqrt{A_1}\pi_k$ in terms of the corresponding mode functions $u_k$, and computing the equation of motion for $u_k$ yields, 
\begin{eqnarray}\label{EOM1}
&&u_k''+ \left(\frac{C_1}{A_1}\frac{k^6}{a^4}+\frac{D_1}{A_1}\frac{k^4}{a^2}+\frac{F_1}{A_1}k^2-\frac{a''}{a}\right) u_k =0\,,
\end{eqnarray}
where the primes denote derivatives with respect to the conformal time $\tau$ such that $f' = \frac{d}{d\tau} f = a\frac{d}{dt} f = a\dot{f}$. Note that in this equation for the positive 
slow varying values of $F_1/A_1$, in the regime where $\frac{F_1}{A_1}k^2$ is the dominant term ($\omega\propto k$) the speed of sound for scalar perturbations is well defined and real, with
\begin{equation} \label{eq:mod_dis}
    c_{{}_S}^2 = \frac{F_1}{A_1}.
\end{equation} 
In the de Sitter limit where $a  = -1/(H\tau)$  
and defining  $y = c_{{}_S}k\tau$ and $\tilde{u}(y)= \sqrt{c_{{}_S} k}\,u_k(\tau)$ to preserve the normalization for Wronskian condition, we can rewrite equation\eqref{EOM1} as
\beq\label{mode-eq-x}
\frac{d^2 \tilde{u}(y)}{d y^2}+\left(1+\alpha_0y^2+\beta_0y^4-\frac{2}{y^2}\right)\tilde{u}(y)=0\,,
\eeq
where
=
\begin{align}
\alpha_0 = \frac{D_1A_1H^2}{F_1^2}, & \text{ \hspace{5mm}and \hspace{5mm}}\beta_0 = \frac{C_1A_1^2H^4}{F_1^3}.
\end{align}
We can solve the mode equation \eqref{mode-eq-x} numerically, taking the positive frequency WKB mode in the infinite past,
\begin{equation}
    \tilde{u}(y\rightarrow-\infty) \simeq \frac{1}{2}\left(-\frac{\pi y}{3}\right)^{1/2}H^{(1)}_{\frac{1}{6}}\left(-\sqrt{\beta_0}\frac{y^3}{3}\right).
\end{equation}
Noting that curvature perturbations $\zeta_k$ are related to $\tilde{u}$ via $ \zeta_k(\tau)=-\frac{H}{a\sqrt{A_1c_{{}_S}k}}\tilde{u}(y) $, the effect of sixth order polynomial dispersion relation of EEFToI on the scalar power spectrum can be obtained. This result can be represented as a modulation factor, $\gamma_{{}_S}$, on the power spectrum expected from the Bunch-Davies vacuum $P_\zeta^{B.D.}$,
\begin{equation}
    P_{{}_\zeta} = \gamma_{{}_S} P_{{}_\zeta}^{B.D.}\,. 
\end{equation} 
Since the dimensionless power spectrum is related to $P_{{}_\zeta}^{B.D}$ as
\begin{eqnarray}\label{eq:powerspectrum}
    \Delta^{B.D.}_\zeta \equiv \frac{k^3}{2\pi^2}P_\zeta^{B.D.}= \frac{1}{4\pi^2}\frac{H^4}{A_1c_{{}_S}}, 
\end{eqnarray}
we also get 
\begin{eqnarray}\label{eq:powerspectrum2}
    \Delta_{{}_{\zeta}} = \frac{\gamma_{{}_S}}{4\pi^2}\frac{ H^4}{A_1c_{{}_S}}. 
\end{eqnarray}
As can be expected, value of $\gamma_{{}_S}$ depends on values of $\alpha_0$ and $\beta_0$. Under the assumption  that the dispersion relation does not become tachyonic before horizon crossing ({\it i.e.} $1+\alpha_0y^2+\beta_0y^4>0$), it was shown in \cite{Ashoorioon:2017toq} that the greatest  enhancements to the Bunch-Davies power spectrum is obtained when $\alpha_0<0$. In particular, with the assumption that the mode never becomes tachyonic while it is inside the horizon, we obtained the maximum enhancement for $\alpha_0 = -0.18$ and $\beta_0 = \alpha_0^2/4$. 
For $\alpha_0<0$ and $\beta_0 \leq \alpha_0^2/4$, the dispersion relation can become tachyonic, allowing for larger enhancements of the power spectrum \cite{Ashoorioon:2018uey,Ashoorioon:2019xqc}. For $\alpha_0,~ \beta_0\geq 0$, the scalar power spectrum get suppressed with respect to its Bunch-Davies counterpart, $\gamma_{{}_S}\lesssim 1$. This observation along with the  fact that for pure quartic positive correction to the dispersion relation, $\gamma_{{}_S}<1$ too \cite{Ashoorioon:2011eg}, drives us toward this conclusion that all positive correction to the dispersion relation has the effect of diminishing the power spectrum, For small values of $0\leq \alpha_0,~ \beta_0\ll 1$, $\gamma_{{}_S}\approx 1$. This agrees with our intuition that if WKB approximation is not violated and the mode changes adiabatically from one regime to another, we do not expect modes to get very excited and they should remain close to the adiabatic vacuum. Figure \eqref{fig:two graphs} illustrates some of these effects for different values of $\alpha_0$ and $\beta_0$.

Now that we have reviewed the main findings of EEFToI at the leve of power spectrum, we can move on to the main goal of this paper, which is computing the third order terms of EEFToI action, bispectrum and finally the non-Gaussianity signatures or additional constraints for EEFToI models. 

\section{Calculation of the Bispectrum and Estimating the non-Gaussianity}\label{section:ng}

In the previous section, we reviewed  the impact of some interesting regions of parameter space in the EEFToI on the power spectrum, where the perturbations start with modified dispersion relation $\omega^2\propto k^6$. The computation of bispectrum and thus the estimation of non-Gaussianity for EEFToI can help us explore more distinctive signatures of these models or further constrain the parameter space of the EEFToI theory. In Fourier space, the bispectrum is related to the vacuum expectation value of the three point function for $\zeta_{\textbf{k}}$ as 
\begin{equation} \label{eq:bispectrum}
    \langle\zeta_{\textbf{k}_1}(t)\zeta_{\textbf{k}_2}(t)\zeta_{\textbf{k}_3}(t)\rangle = (2\pi)^3\delta^3(\textbf{k}_1+\textbf{k}_2+\textbf{k}_3)B(k_1,k_2,k_3)\,, 
\end{equation}
and the three point function in the interaction picture can be computed as
\begin{equation}\label{eq:interaction-threepoint}
    \langle\zeta_{\textbf{k}_1}(t)\zeta_{\textbf{k}_2}(t)\zeta_{\textbf{k}_3}(t)\rangle=-i\int^t_{t_{in}}dt'\;\langle[\zeta_{\textbf{k}_1}(t)\zeta_{\textbf{k}_2}(t)\zeta_{\textbf{k}_3}(t),H_{int}(t')]\rangle\,,
\end{equation} where in these calculations, $\pi_{\textbf{k}}$ can be substituted for $\zeta_{\textbf{k}}$ using the relation $\zeta_{\textbf{k}} = -H\pi_{\textbf{k}}$ and $t_{in}$ is the initial time. The interaction Hamiltonian is given by $H_{int} = -L_{int}$ which includes terms beyond quadratic order in the action\cite{Cheung:2007st}. 

\par Before we do a robust computation of the three point functions, let us perform a simple analysis to see how the interaction terms in the action act under the energy scaling when the dispersion relation is governed by $\omega^2\propto k^6$.
The kinetic term in the action \eqref{KEFT} in real space has the following form for $\pi$ 
\begin{equation}
    S=\int dt\;d^3x\; [\frac{1}{2}A_1\dot{\pi}^2]. 
\end{equation}
Rescaling energy by a factor of $\Sigma$, $E\rightarrow \Sigma E$, means that $t\rightarrow \Sigma^{-1}t$, which given the dispersion relation $\omega^2\propto k^6$, implies $ x \rightarrow \Sigma^{-1/3} x$. 
Assuming $A_1$ variation in time is negligible (due to generalized slow-roll conditions) the dimensional scaling implies that for the kinetic term to remain invariant, $\pi$ must remain scale-invariant as well, $\pi\rightarrow\pi$. 

\par We now check the scaling of a generic term in action, $\int dt~d^3x \;\mathcal{M}_{m,s,p}$ where $m$ denotes the number of {\it time} derivatives, $s$ number of {\it spatial} derivatives and $p$  the overall powers of $\pi$. Since $x\rightarrow \Sigma^{-1/3}x$, the spatial derivative goes as $\partial_i \rightarrow \Sigma^{1/3}\partial_i$ so the contribution to action scales as 
\begin{equation}
    \mathcal{M}_{m,s,p}\;dt\;dx^3\rightarrow \Sigma^{-2+m+s/3}\mathcal{M}_{m,s,p}\;dt\;d^3x.
\end{equation}
When $m+s/3 < 2$, this term diverges in the low energy limit of $\Sigma\rightarrow 0$ and becomes relevant. Note the powers of $\pi$ cannot be more than the number of derivatives, as the time diffeomorphism invariance of the action does not allow Goldstone boson to be produced without derivatives. This gives us the relations $p\leq s+m$.  Also, $p=1$ corresponds to linear terms, which cancel out for on-shell solutions. All possible relevant operators thus satisfy the following constraints,
\begin{eqnarray}
   && 2 \leq p \leq  s+m\,, \\ && m+s/3 <  2\; .
\end{eqnarray}
Furthermore, the spatial diffeomorphism invariance enforces $s$ to be even. Taking into account all of these constraints, the only possible combinations of $(m,s,p)$ that remain are $(0,2,2)$, $(1,2,2)$, $(0,4,2)$, $(0,4,3)$, $(1,2,3)$ and $(0,4,4)$. This implies the number of relevant operators is finite and excludes the existence of any relevant operators beyond quartic terms as well\footnote{This exercise can be done for any $\omega^2\propto k^{2n}$ and there are always finite number of relevant operators.}. Also note that $\mathcal{M}_{1,2,2} \propto \partial\dot\pi\partial_i\pi = \frac{1}{2}\frac{\partial}{\partial t}~(\partial_i\pi)^2$ term becomes $\mathcal{M}_{0,2,2} \propto (\partial_i\pi)^2$ after integrating by parts, which can be absorbed into the canonical kinetic term. The scaling power of $\mathcal{M}_{0,2,2} \propto (\partial_i\pi)^2$ and $\mathcal{M}_{0,4,2}\propto(\partial_i^2\pi)^2$ which modify the dispersion relation, as we had pointed out earlier are respectively $-4/3$ and $-2/3$. The scaling power of $\mathcal{M}_{1,2,3} \propto \dot{\pi} (\partial_i\pi)^2$ is $-1/3$ while both $\mathcal{M}_{0,4,2} \propto \partial_j^2\pi(\partial_i\pi)^2$ and $\mathcal{M}_{0,4,4} \propto (\partial_i\pi)^4$ scale as $-2/3$ as well. 
The term $\mathcal{M}_{0,4,4} \propto (\partial_i\pi)^4$ contributes to trispectrum which we will not discuss in this paper\footnote{See \cite{Bartolo:2010di} for calculation of trispectrum in EFToI, with at most quartic dispersion relation.}. 
Therefore, to calculate the strongest contributions to bispectrum we only need to consider operators $(0,4,3)$ and $(1,2,3)$, which are 
\begin{equation}
    (\partial_i\pi)^2\dot\pi
\end{equation} 
\begin{equation}
    \partial^2_j\pi(\partial_i\pi)^2.
\end{equation} 
Note that both of these terms are also produced in \eqref{Lpi} from the terms included in the EFToI as well:
\begin{equation}
    \frac{1}{2}M_2^4(g^{00}+1)^2\xrightarrow{\{\mathcal{O}(\pi^3)\}} -2M_2^4\frac{\dot\pi(\partial_i \pi)^2}{a^2}
\end{equation} and 
\begin{eqnarray}
    &&-\frac{1}{2}\bar{M}_1^3(g^{00}+1)\delta K^\mu_\mu-\frac{\bar M_2^{2}}{2}\, (\delta K^\mu_{\ \mu})^2
-\frac{\bar M_3^{2}}{2}\, \delta K^\mu_{\ \nu}\,\delta K^\nu_{\ \mu}\nonumber \\ &&\xrightarrow{\{\mathcal{O}(\pi^3)\}} -\frac{1}{2}\left(\bar{M}_1^3+\bar{M}_2^2H+2\bar{M}_3^2 H\right)\frac{\partial_i^2\pi(\partial_j\pi)^2}{a^4}.
\end{eqnarray} 
These cubic terms get additional contributions in the extended theory. The detailed calculation can be found in the appendix \eqref{AppendixA} and the corrections after setting $\delta_3 = -3\delta_4$ are the following 
\begin{align}
    -\frac{\delta_3}{2}\nabla^\mu\delta& K_{\nu\mu}\nabla^\nu\delta{K^\sigma}_\sigma
    -\frac{\delta_4}{2}\nabla_\mu\delta{K^\mu}_\nu\nabla_\gamma\delta K^{\gamma\nu}
    \nonumber\\
    &\xrightarrow{\{\mathcal{O}(\pi^3)\}}\left(\frac{\delta_3H^3}{3}\right)\frac{ \partial_i^2\pi(\partial_j\pi)^2 }{a^4}
    -\delta_3 H^4\frac{\dot\pi (\partial_i\pi)^2}{a^2}
\end{align} and 

\begin{dmath}
\frac{\bar{M}_4}{2}\nabla^\mu \delta g^{00}\nabla^\nu\delta K_{\mu\nu}\xrightarrow{\{\mathcal{O}(\pi^3)\}} 9 \bar{M}_4 H^3  \frac{\dot\pi(\partial\pi)^2 }{a^2} + \frac{1}{2}\bar{M}_4H^2\frac{ \partial_i^2\pi(\partial_j\pi)^2 }{a^4}
\end{dmath}
In the limit where the parameters of EFToI and EEFToI satisfy 
 \begin{equation}
     -\delta_3H^4 +9\bar{M}_4H^3-2M_2^4 = 0,
 \end{equation}
 and
 \begin{equation}
     \frac{1}{3}\delta_3H^3 -  \frac{1}{2}\left(\bar{M}_1^3+\bar{M}_2^2H+2\bar{M}_3^2H-\bar{M}_4H^2\right) = 0, 
 \end{equation}
these terms will not lead to any strong coupling or non-Gaussianities. However, aside from these particular corners of parameter space, we show below that we do not need to worry about the impact of these terms. We start by making a crude estimation of the contribution of these terms by directly comparing them to quadratic terms as they appear in action. Around freezing we have $\omega \sim H$ and $k \sim H/c_{{}_S}$ and approximating  $\dot\pi\sim\omega\pi$, the ratio of contributions of the cubic term $(\partial_i\pi)^2\dot\pi$ to the quadratic term can be estimated as 

\begin{eqnarray}
    \frac{\mathcal L_{\dot{\pi} (\partial\pi)^2}}{\mathcal L_2} &\sim & (-\delta_3H^4 +9\bar{M}_4H^3-2M_2^4)\frac{\dot{\pi} (\partial\pi)^2}{\frac{A_1}{2}{\dot{\pi} }^2} \xrightarrow{\pi_k} 2(-\delta_3H^4 +9\bar{M}_4H^3-2M_2^4)\frac{\omega k^2\pi_k^3}{A_1\omega^2\pi
    _k^2}\nonumber \\
    &\sim& 2(-\delta_3H^4 +9\bar{M}_4H^3-2M_2^4)\frac{H\pi(\frac{H^2}{c_{{}_S}^2}\pi^2)}{A_1H^2\pi^2}\sim 2(-\delta_3H^4 +9\bar{M}_4H^3-2M_2^4)\frac{H}{A_1c_{{}_S}^2}\pi\nonumber,\\&\sim & 2\frac{(-\delta_3H^4 +9\bar{M}_4H^3-2M_2^4)}{A_1c_{{}_S}^2}\zeta\sim 8\pi^2\left(- \delta_3+9\frac{\bar{M}_4}{H}-2\frac{ M_2^4}{H^4}\right) \frac{\Delta_{{}_{\zeta}}}{c_{{}_S} \gamma_{{}_S}}\zeta
\end{eqnarray} 
and similarly for $\partial^2_j\pi(\partial_i\pi)^2$, we get
\begin{eqnarray}
    \frac{\mathcal L_{(\partial\pi)^2\partial^2\pi}}{\mathcal L_2}&\sim& \left(\frac{2}{3}\delta_3H^3+\bar{M}_4H^2 -  \left(\bar{M}_1^3+\bar{M}_2^2H+2\bar{M}_3^2H\right)\right) \frac{\left(\frac{H}{c_{{}_S}}\pi\right)^2\left(\frac{H}{c_{{}_S}}\right)^2\pi}{A_1H^2\pi^2}\nonumber\\
    &\sim& \left(\frac{2}{3}\delta_3H^3+\bar{M}_4H^2 -  \bar{M}_1^3-\bar{M}_2^2H-2\bar{M}_3^2H\right)\frac{H^2}{A_1c_{{}_S}^4}\pi \nonumber\\
    &\sim& \left(\frac{2}{3}\delta_3H^3 +\bar{M}_4H^2-  \bar{M}_1^3-\bar{M}_2^2H-2\bar{M}_3^2H\right)\frac{H}{A_1c_{{}_S}^4}\zeta \nonumber\\&\sim& 4 \pi^2\left(\frac{2}{3}\delta_3 -  \frac{\bar{M}_1^3}{H^3}-\frac{\bar{M}_2^2}{H^2}-2\frac{\bar{M}_3^2}{H^2}+\frac{\bar{M}_4}{H}\right)\frac{\Delta_{{}_{\zeta}}}{c_{{}_S}^3 \gamma_{{}_S}}\zeta \nonumber \\
\end{eqnarray} 
The perturbation theory breaks down when the above ratios become of order one but given that amplitude of curvature perturbation is of order $10^{-5}$ implying $\zeta\Delta_{{}_{\zeta}}\sim 10^{-15}$, then it is reasonable to expect that there is a large viable window of parameter space where strong coupling is avoided. However, as we discuss later CMB observations set a much stronger constraints on this ratios due to the limits they set on the non-Gaussianity. 
\par To see the restrictions on the size of non-Gaussianity from the CMB observations, we go back to the more rigorous calculation of bispectrum and the the contribution of the cubic terms through the interaction Hamiltonian and equations \eqref{eq:bispectrum} and \eqref{eq:interaction-threepoint}. Note that the Lagrangian and the action are written in terms of $\pi$, and we are interested in the comoving curvature perturbation, $\zeta = -H\pi$, and its bispectrum $<\zeta^3>$. Therefore, we transform the terms $H_{int}(\pi) \rightarrow H_{int}(\zeta)$ and also for convenience switch to conformal time $\tau$ instead of $t$. As we pointed out earlier, the most dangerous contributions from $H_{int}$ or $L_{int}$ are coming from terms $\dot\pi(\partial_i\pi)^2$ and $\partial_i^2\pi(\partial_j\pi)^2$. In terms of $\zeta$, these terms become 
\begin{eqnarray}
    \dot{\pi}(\partial\pi)^2\rightarrow  -\frac{1}{aH^3}\zeta'(\partial\zeta)^2\\
    \partial^2\pi(\partial\pi)^2\rightarrow-\frac{1}{H^3}\partial^2\zeta(\partial\zeta)^2.
    \end{eqnarray}
Next, similar to calculating power spectrum, $\zeta_k$ is written as a quantum field operator in terms of the creation and annihilation operators $a_{\textbf{k}}^\dagger$, $a_{-\textbf{k}}$ satisfying the canonical commutation relations $[a_{\textbf{k}},a_{\textbf{k}}^\dagger] = (2\pi)^3\delta^3(\textbf{k}-\textbf{k}')$ and the mode functions $f_k(\tau)$:
\begin{equation} \label{mfunction}
    \zeta(\textbf{k},\tau) = a_{-\textbf{k}} f_k(\tau) + a_{\textbf{k}}^\dagger f^*_k(\tau).
\end{equation} 
 In practice, we carried out the calculations numerically in terms of the modes functions $\tilde{u}(c_{{}_S} k \tau) = -\frac{ \sqrt{c_{{}_S} k A_1}a}{H} f_k(\tau)$ which we had computed previously from quadratic action and performed the integration in terms of the dimensionless variable $x_i=c_{{}_S}k_i\tau$ \footnote{See appendix \eqref{numericalcalculations} for more details.}. Under these changes of variables, contributions from the dominant cubic interaction terms to the three point function were found to be as follows,
 
\indent 1. contributions from the $\dot\pi(\partial\pi)^2$ term to  $\left<\zeta(\textbf{k}_1)\zeta(\textbf{k}_2)\zeta(\textbf{k}_3)\right>$ is:
\begin{align}\label{eq:pdcontribution}
   \left<\zeta(\textbf{k}_1)\zeta(\textbf{k}_2)\zeta(\textbf{k}_3)\right>_{\dot\pi(\partial\pi)^2}=&(2\pi)^3\delta^3(\textbf{k}_3+\textbf{k}_2+\textbf{k}_1)\frac{H^{12}}{c_{{}_S}^8 A_1^3} \left[-\delta_3+9\frac{\bar{M}_4}{H}-2\left(\frac{M_2}{H}\right)^4\right]  \nonumber\\& \left [i\frac{(1 - \lambda^2 - \theta^2)x_f^3}{k_1^6\lambda\theta}\left\{\int_{-\infty}^{x_{f}}dx_1'\; (x_1')^2 \right. \right. 
  \tilde{u}(x_f){\tilde{u}^*}_{,x}(x_1')\nonumber\\&\tilde{u}(\lambda \;x_f){\tilde{u}^*}(\lambda \; x_1')\tilde{u}(\theta\; x_f){\tilde{u}^*}(\theta\; x_1') \nonumber\\&
  +\int_{-\infty}^{x_f} dx_1'\; x_1'\;\tilde{u}( x_f){\tilde{u}^*}(x_1')\tilde{u}(\lambda \;x_f){\tilde{u}^*}(\lambda \;x_1') \nonumber\\
   & 
  \tilde{u}(\theta\; x_f){\tilde{u}^*}(\theta\; x_1')+ {\rm c.c.}\bigg\}\bigg]+ {\rm cycl.}
\end{align}
where we are defining parameters $\lambda \equiv \frac{k_2}{k_1}$ and $\theta \equiv \frac{k_3}{k_1}$ that determine the shape of the triangles formed by $\textbf{k}_1,~ \textbf{k}_2 $ and $\textbf{k}_3$.

\indent 2. contributions from the $\partial^2\pi(\partial\pi)^2$ term to  $\left<\zeta(\textbf{k}_1)\zeta(\textbf{k}_2)\zeta(\textbf{k}_3)\right>$ is: 
\indent \begin{align}\label{eq:pppcontribution}
    \left<\zeta(\textbf{k}_1)\zeta(\textbf{k}_2)\zeta(\textbf{k}_3)\right>_{\partial^2\pi(\partial\pi)^2}=&(2\pi)^3\delta^3(\textbf{k}_3+\textbf{k}_2+\textbf{k}_1)\nonumber\\&\frac{H^{12}}{c_{{}_S}^{10} A_1^3}\left[\frac{1}{3}\delta_3-  \frac{1}{2}\left(\left(\frac{\bar{M}_1}{H}\right)^3+\left(\frac{\bar{M}_2}{H}\right)^2+2\left(\frac{\bar{M}_3}{H}\right)^2-\frac{\bar{M}_4}{H}\right)\right]\nonumber\\
   & \left[i \frac{(1 - \lambda^2 - \theta^2)x_{f}^3}{k_1^6\lambda\theta}\int_{-\infty}^{x_{f}} dx_1' (x_1')^3\; \tilde{u}(x_{f}){\tilde{u}^*}(x_1')\tilde{u}(\lambda x_{f})\right.\nonumber\\& {\tilde{u}^*}(\lambda x_1')\tilde{u}(\theta x_{f}){\tilde{u}^*}(\theta x_1')+ {\rm c.c.}\bigg]+ {\rm cycl.}
\end{align}

\begin{figure}[ht] 
     \centering
     \begin{subfigure}[b]{0.7\textwidth}
         \centering
         \includegraphics[width=\textwidth]{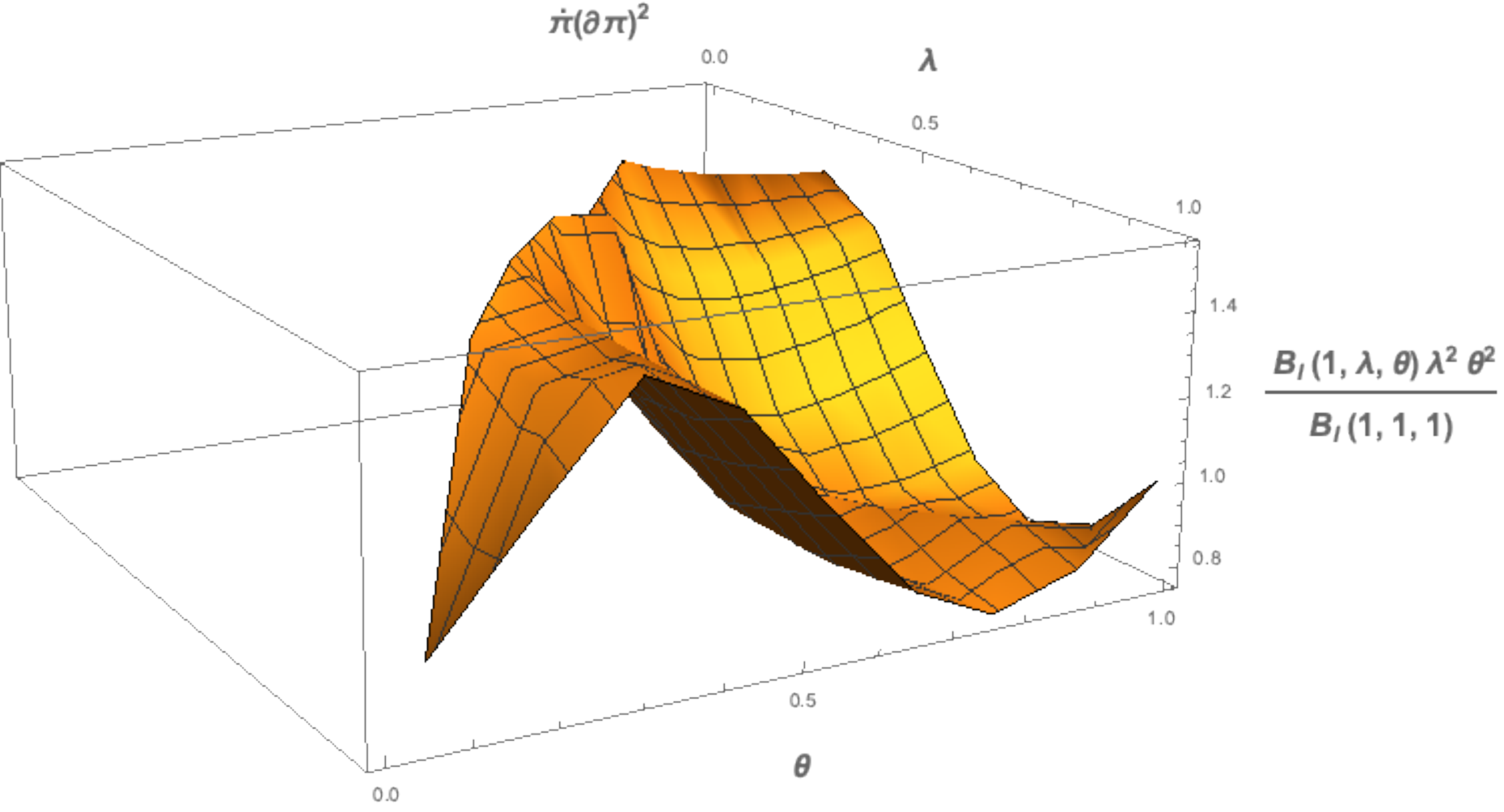}
         \caption{The shape of bispectrum for operator  $\dot\pi(\partial\pi)^2$ }
     \end{subfigure}
     \begin{subfigure}[b]{0.7\textwidth}
         \centering
         \vspace{0.5cm}
         \includegraphics[width=\textwidth]{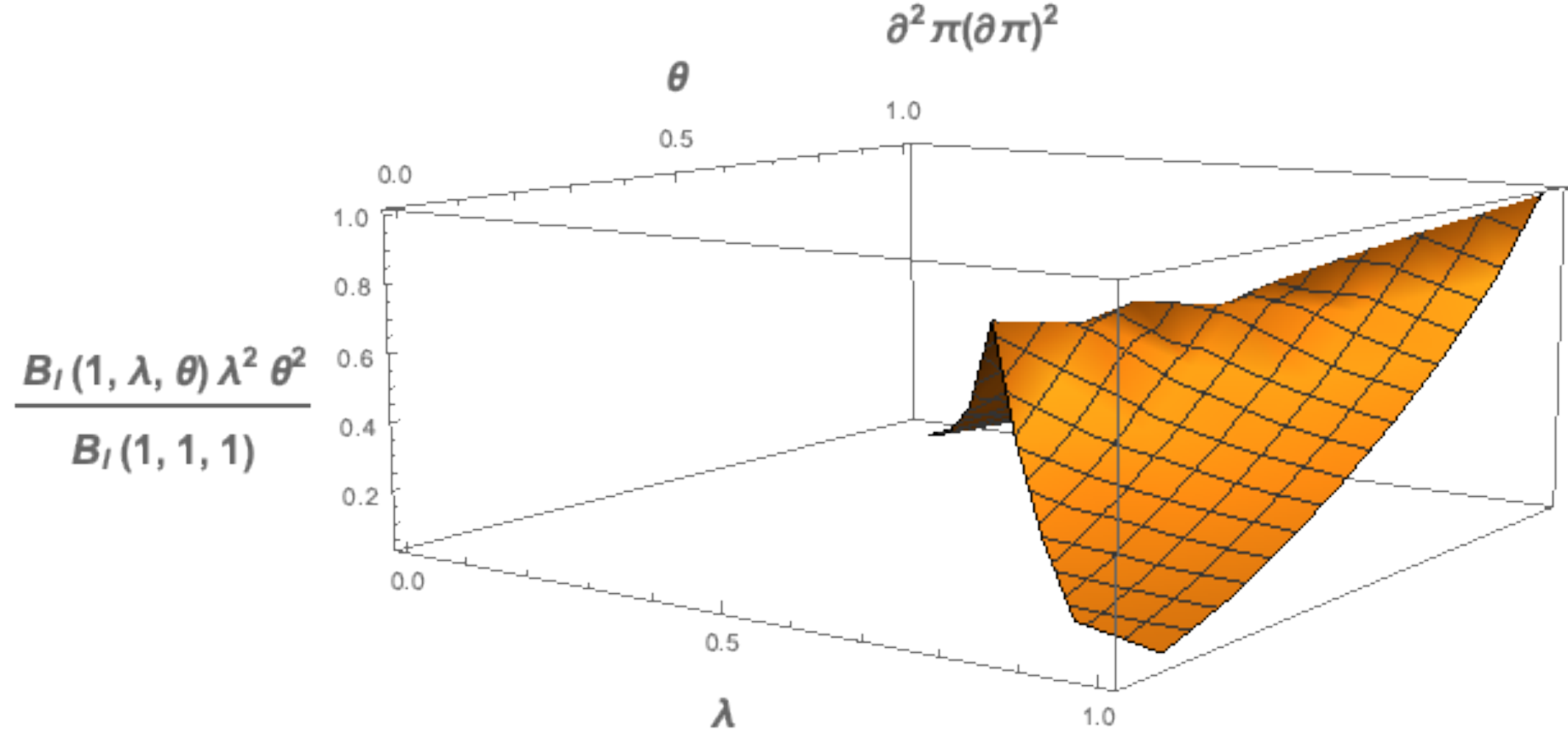}
          \caption{The shape of bispectrum for operator  $\dot\pi(\partial\pi)^2$}
     \end{subfigure}
     \caption{Calculation of the quantity $B_I(k_1,k_2,k_3)k_1^2k_2^2k_3^2$, evaluated at $k_1=1$ and normalized over $B_I(1,1,1)$ for various triangle shapes. This result corresponds to value of $\alpha_0=-0.18$ and $\beta_0=\alpha_0^2/4 $.} 
    \label{fig:BI}
\end{figure}

The above derivations allows to numerically evaluate bispectrum for different shapes of triangles at different scales and for given values of parameters. To simplify connecting these expressions to our numerical calculations, we break down their contribution to bispectrum from each interaction, $I$, as 
\begin{equation}\label{BIbreak}
(4\pi^2 \Delta^{B.D.}_{\zeta})^2 ~ \tilde{q}_I~\left[\frac{1}{k_1^6}G_I(x_f,\lambda,\theta)+{\rm sym}\right]\longrightarrow B_{I}(k_1,k_2,k_3)\end{equation}
where $(4\pi^2 \Delta^{B.D.}_{\zeta})^2=\frac{H^8}{ A_1^2 c_{{}_S}^2}$
from \eqref{eq:powerspectrum} and
\begin{align}\label{defq}
    \tilde{q}_{\partial^2\pi(\partial\pi)^2}&\equiv \frac{H^{4}}{c_{{}_S}^6 A_1} \left[-\delta_3+9\frac{\bar{M}_4}{H}-2\left(\frac{M_2}{H}\right)^4\right]\,, \\
    \tilde{q}_{\dot\pi(\partial\pi)^2}&\equiv\frac{H^{4}}{c_{{}_S}^6 A_1}\left[\frac{1}{3}\delta_3-  \frac{1}{2}\left(\left(\frac{\bar{M}_1}{H}\right)^3+\left(\frac{\bar{M}_2}{H}\right)^2+2\left(\frac{\bar{M}_3}{H}\right)^2-\frac{\bar{M}_4}{H}\right)\right]\,.
\end{align}
We can now numerically calculate contributions to bispectrum for both operators at various triangle shapes. Figure \eqref{fig:BI} displays one of our results for $B_I(k_1,k_2,k_3)k_1^2k_2^2k_3^2$, evaluated at $k_1=1$ and normalized over $B_I(1,1,1)$, for various triangle shapes. 
As we see the bispectrum seems to be enhanced for triangle configurations other than just folded \cite{Holman:2007na,Ashoorioon:2010xg} or squeezed type \cite{Agullo:2010ws,Ashoorioon:2013eia}. This generically agrees to what one expects from non-Bunch-Davies vacuum initial conditions although each manifestation leads to distinct bispectrum enhancement for different shapes and makes it possible to distinguish them from each other \footnote{The calculation of bispectrum due to these operators for EFToI models with $\omega^2\propto k^4$ have also been shown to have non-analytical forms and result in very different shapes from the folded and local kind\cite{Senatore:2009gt, Bartolo:2010bj}.}. 
Now to make the connection with observations, we need to turn observational constraints on bispectrum, into bounds on parameters of our models by substituting for different mass and coupling parameters in $\tilde{q}_i$ and other variables determining $B_I$. We will further explore these bounds and discuss our numerical calculations in the next section. 

\section{Numerical Results and Constraints on the Models}\label{sec:eeftoiresult}
We know provide  more details on our numerical calculations and how we compared them with observtions in order to interpret the constraints on our model.

To quantify the strength of non-gaussianity due to terms discussed in last section and connect them to observations, it is conventional to express the result in terms of the quantities known as $f_{{}_{\rm NL}}$ in the literature. 

Historically, $f_{{}_{\rm NL}}$ was first defined to characterise the {\it local} non-Gaussianity in CMB data \cite{Komatsu_2001}. In that case, one assumes the non-Gaussian corrections for curvature perturbation can be expressed locally as 
\begin{equation}\label{fnllocal}
    \zeta = \zeta_{{}_{\rm L}} + \frac{3}{5}f_{{}_{\rm NL}}\zeta_{{}_{\rm L}},
\end{equation} where $\zeta_{\rm L}$ is the gaussian scalar perturbation. Now under this assumption as well as near scale invariance (i.e. $n_{{}_{\rm S}}-1\ll 1$), one can obtain the following relationship between the three point function in $\textbf{k}$ space and the dimensionless power spectrum of $\zeta$ \footnote{See appendix \eqref{define FNL} for details.}, 
\begin{equation}\label{localNG3point}
     \left<\zeta(\textbf{k}_1)\zeta(\textbf{k}_2)\zeta(\textbf{k}_3)\right>=(2\pi)^3\delta^3(\textbf{k}_1+\textbf{k}_2+\textbf{k}_3)[(2\pi)^4\Delta_{{}_{\zeta}}^2] ~f_{{}_{\rm NL}}^{{}^{\rm local}} \times \left\{ \frac{3}{10}~\bigg(\frac{k_1^3+k_2^3+k_3^3}{k_1^3k_2^3k_3^3}\bigg)\right\}. 
\end{equation}
Therefore, assuming a local form bispectrum and comparing \eqref{eq:bispectrum} and \eqref{localNG3point}, one can construct a dimensionless quantity directly from bispectrum itself consistent with the definition in \eqref{fnllocal},
\begin{equation}
    f_{{}_{\rm NL}}^{{}^{\rm local}} =\left(\frac{10}{3}\frac{\prod k_i^3}{\sum k_i^3}\right)\times  \frac{1}{(2\pi)^4\Delta_{{}_{\zeta}}^2}B(\textbf{k}_1,\textbf{k}_2,\textbf{k}_3). 
\end{equation}
In practice, even if the theory produces exact local form bispectrum, the data is going to be contaminated with noise and other secondary effects so direct application of above formula to data would lead to extracting different values of $f_{{}_{\rm NL}}^{{}^{\rm local}}$ for different triangles in momentum spaces,
\begin{equation}
    f_{{}_{\rm NL}}^{{}^{\rm local}}(\textbf{k}_1,\textbf{k}_2,\textbf{k}_3) = \left(\frac{10}{3}\frac{\prod k_i^3}{\sum k_i^3}\right)\times  \frac{1}{(2\pi)^4\Delta_{{}_{\zeta}}^2}B(\textbf{k}_1,\textbf{k}_2,\textbf{k}_3).
\end{equation}
Therefore, finding the observed value of $f_{{}_{\rm NL}}^{{}^{\rm local}}$ is a more subtle task of optimisation of the fitting between the template in \eqref{localNG3point} and all the data points and it involves statistical schemes beyond the scope of this work. It can easily be seen that the particular momentum dependence of \eqref{localNG3point} diverges in squeezes limits where one of the modes $k_i\to 0$, so this template is suited for fitting models where contributions from those corners are enhanced. 
\par However, in general the three point function for different interactions and in different models does not necessarily lead to $\textbf{k}$-dependence presented in \eqref{localNG3point} and it can have crests and troughs at different locations. Therefore, over the years there have been different templates developed to best extract possible non-gaussianities in data\footnote{Readers are referred to \cite{planckcollaboration2019planck} for details on different templates fitting for different models.} in order to test different types of models. The next famous example are non-canonical single scalar field inflationary models  \cite{Armendariz-Picon:1999hyi,Alishahiha:2004eh}  that usually peak at the limits of equilateral shape triangles \cite{Chen:2006nt} and can be approximated by 
\begin{eqnarray}\label{equilNG3point}
     \left<\zeta(\textbf{k}_1)\zeta(\textbf{k}_2)\zeta(\textbf{k}_3)\right>^{{}^{\rm equil.}}&=&(2\pi)^3\delta^3(\textbf{k}_1+\textbf{k}_2+\textbf{k}_3)[(2\pi)^4\Delta_{{}_{\zeta}}^2] ~f_{{}_{\rm NL}}^{{}^{\rm equil.}}\times \bigg(-\frac{9}{10}\bigg) \nonumber\\&&\left\{ \left[\frac{1}{k_1^3k_2^3}+\text{cycl.}\right]+\frac{2}{k_1^2k_2^2k_3^2}-\left[\frac{1}{k_1k_2^2k_3^3}+(5~ \text{perm.})\right]\right\}.  
\end{eqnarray}
For agnostic bispectra interpretation one can perform a more sophisticated mathematical approach of projecting the observed bispectra on different known and independent templates by defining an inner product between bispectrum and these templates \cite{Senatore:2009gt, Meerburg:2009ys} and then use joint estimators to simultaneously fit for several types of non-gaussinities. Of course, given that these functions do not form a complete basis for all the possible models they may not necessary capture all aspects of non-gaussianity. 
Non-Bunch-Davies bispectra have been found to get enhanced for both the flattened and squeezed shapes \cite{Chen:2006nt,Holman:2007na,Agullo:2010ws,Ashoorioon:2010xg,Ashoorioon:2013eia}. Planck collaboration \cite{planckcollaboration2019planck} has used variety of these templates as model estimators in order to test non-Bunch-Davies (NBD) vacuum initial conditions. 
For example, one of the simple flatten triangles shaped enhancements that can rise due to NBD models \cite{Chen:2006nt,Holman:2007na} is 
\begin{eqnarray}\label{enfoldedNG3point}
     \left<\zeta(\textbf{k}_1)\zeta(\textbf{k}_2)\zeta(\textbf{k}_3)\right>^{{}^{\rm NBD}}&=&(2\pi)^3\delta^3(\textbf{k}_1+\textbf{k}_2+\textbf{k}_3)[(2\pi)^4\Delta_{{}_{\zeta}}^2] ~f_{{}_{\rm NL}}^{^{\rm flat}} \nonumber\\&&\times \left\{ \frac{3}{10}~\bigg(\frac{1}{\prod k_i}\bigg)\left[\frac{1}{(k_2+k_3-k_1)^3}+\text{cycl.}\right]\right\}+\dots\,, 
\end{eqnarray}
which diverges as $k_1\to k_2+k_3$. 
Another template that has been used for testing EFToI models is the orthogonal template defined as
\begin{eqnarray}\label{orthoNG3point}
     \left<\zeta(\textbf{k}_1)\zeta(\textbf{k}_2)\zeta(\textbf{k}_3)\right>^{{}^{\rm ortho.}}&=&(2\pi)^3\delta^3(\textbf{k}_1+\textbf{k}_2+\textbf{k}_3)[(2\pi)^4\Delta_{{}_{\zeta}}^2] ~f_{{}_{\rm NL}}^{{}^{\rm ortho}} \times \bigg(-\frac{9}{10}\bigg) \nonumber\\&&\left\{ \left[\frac{3}{k_1^3k_2^3}+\text{cycl.}\right]+\frac{8}{k_1^2k_2^2k_3^2}-\left[\frac{3}{k_1k_2^2k_3^3}+(5\,\text{perm.})\right]\right\}. 
\end{eqnarray}
For our purposes here, since we want to make only order of magnitude estimations and since  $f_{{}_{\rm NL}}$ constraints from observations are only reported for these known template, we will make a crude estimation by making a linear fit to three templates of local, equilateral and orthogonal templates and then compare our result to observational constraints reported for these shapes. However, as figure \eqref{fig:BI} indicates the shape of bispectrum can not be described through just linear combination of these templates and there is some residual non-gausianities left. A more robust way of constraining our model which is beyond the scope of this work, would be direct comparison of our numerical template with the data. 
However, here after substituting for $B_I(\textbf{k}_1,\textbf{k}_2,\textbf{k}_3)$ from \eqref{BIbreak} and that $\Delta_{{}_{\zeta}}= \gamma_{{}_S} \Delta^{B.D}_\zeta$, we approximate different contributions to $f_{{}_{NL}}^{{}^{\rm local}}, ~f_{{}_{NL}}^{{}^{\rm equil.}}$ and $f_{{}_{NL}}^{{}^{\rm ortho.}}$ in terms of $G_I$ and $\tilde{q}_I$ as following linear combination
\begin{equation}
    \frac{\tilde{q}_I~}{\gamma_{{}_S}^2}\left[\frac{1}{k_1^6}G_I(x_f,\lambda,\theta)+ \text{cycl}\right]= f_{{}_{\rm NL}}^{{}^{(I)\rm local}}~T^{{}^{\rm local}}+ f_{{}_{\rm NL}}^{{}^{(I)\rm equil.}}~T^{{}^{\rm equil.}}+f_{{}_{NL}}^{{}^{(I)\rm ortho.}}~T^{{}^{\rm ortho.}}. 
\end{equation} 
The template functions $T^{(\dots)}(k_1,k_2,k_3)$ are the templates mentioned in \eqref{localNG3point}, \eqref{equilNG3point} and \eqref{orthoNG3point}: 
\begin{eqnarray}
  T^{{}^{\rm local}}(k_1,k_2,k_3) &=& \frac{3}{10}~\left\{\frac{k_1^3+k_2^3+k_3^3}{k_1^3k_2^3k_3^3}\right\} \nonumber \\
  T^{{}^{\rm equil.}}(k_1,k_2,k_3) &=&-\frac{9}{10}\left\{ \left[\frac{1}{k_1^3k_2^3}+\text{cycl.}\right]+\frac{2}{k_1^2k_2^2k_3^2}-\left[\frac{1}{k_1k_2^2k_3^3}+(5\, \text{perm.})\right]\right\} \\
T^{{}^{\rm ortho.}}(k_1,k_2,k_3) &=& -\frac{9}{10}\left\{ \left[\frac{3}{k_1^3k_2^3}+\text{cycl.}\right]+\frac{8}{k_1^2k_2^2k_3^2}-\left[\frac{3}{k_1k_2^2k_3^3}+(5\, \text{perm.})\right]\right\}.
\end{eqnarray}
Given that $k_2/k_1 = \lambda$ and $k_3/k_1 = \theta$ the problem reduces to finding the best fit for the coefficients $\frac{f_{{}_{\rm NL}}^{(I)\dots}\gamma_{{}_S}^2}{\tilde{q}_I~}$ in this relation
\begin{eqnarray}
  \left[\frac{1}{k_1^6}G_I(x_f,\lambda,\theta)+ \text{cycl.}\right]_{k_1=1}\approx &&  \frac{f_{{}_{\rm NL}}^{{}^{(I)\rm local}}\gamma_{{}_S}^2}{\tilde{q}_I~}T^{{}^{\rm local}}(1,\lambda,\theta)+\frac{f_{{}_{\rm NL}}^{{}^{ (I) \rm equil.}}\gamma_{{}_S}^2}{\tilde{q}_I~}T^{{}^{\rm equil}}(1,\lambda,\theta)\nonumber\\&&+\frac{f_{{}_{\rm NL}}^{{}^{(I)\rm ortho.}}\gamma_{{}_S}^2}{\tilde{q}_I~}T^{{}^{\rm ortho.}}(1,\lambda,\theta).
\end{eqnarray} 
\par Next, we provide a brief summary of our numerical scheme for readers. Note that as it was described in section \eqref{section:EFT_Review}, we had computed power spectrum which itself involved numerical computation of the canonical mode function $\tilde{u}(y)$ where $y= c_{{}_S}k\tau$. This computation was carried out for different values of $\alpha_0$ and $\beta_0$ that themselves are functions of the parameters in the EEFToI action. We had found empirically that in the regime that the dispersion relation does not become tachyonic inside the horizon, the amplification to the power spectrum $\gamma_{{}_S}$ was the largest (see Figure \eqref{fig:two graphs}) in the following region 
\begin{equation}
    -0.5\leq \alpha_0\leq-0.1\text{\hspace{5mm}and\hspace{5mm}}0\leq\frac{\alpha_0^2}{4}\leq\beta_0\leq\frac{\alpha_0^2}{3}.
\end{equation}
In fact, within that region the enhancement was found to be the greatest ($\gamma_{{}_S}\geq 14.6$) when setting $\alpha_0=-0.18$ and $\beta_0=\alpha_0^2/4$. From physics point of view, a more interesting applications of the EEFToI and EFToI is when they result in larger values of $\gamma_{{}_S}$ or significant deviations from the Bunch Davis initial conditions. Therefore, it is also more interesting to study the impact of observational constraints on bispectrum in this particular region. 
To do that, numerically computed values of $\tilde{u}(y)$ for different values of $\alpha_0$ and $\beta_0$ were substituted in the equations \eqref{eq:pdcontribution} and \eqref{eq:pppcontribution} and then numerically integrated to obtain the contributions to bispectrum.
\par In order to set the initial conditions at $x\to -\infty$, we chose the value of $x_{initial}$ to be where $\beta_0 x^4$ is far greater than $\alpha_0x^2$ ($x_{initial} \sim -50 |\alpha_0/\beta_0|^{1/2}$). This guarantees that the initial condition is set in the regime where $\omega^2 \propto k^6$ dominates the dispersion relation. For the upper limit of the integration, we choose $x_{f} = -0.00001$. This corresponds to setting different values of $\tau_f$ for different $k_i$, but ensures all the modes that were considered had crossed the freezing radius at that point. We found the result of the computation was not very sensitive to changing $x_{initial}$ further into the past or $x_{f}$ into the present. 
The triangle shapes that were explored corresponded to values of $0.2\leq\lambda\leq1$ and $0.2\leq\theta\leq1$. The lower bounds for $\lambda$ and $\theta$ should in principle either be set by ratio of the observed modes in CMB ($\sim 10^3$) or theoretically by validity of the effective field theory regime. Note, the extreme squeezed limit $\theta \to 0$ or $\lambda \to 0$ can push the initial conditions for two of the modes into ultraviolet scales or the other into infrared. Therefore, one needs to ensure all of the mode functions such as $\tilde{u}(\lambda x)$ also initiate in the same regime. In practice, setting $0.2$ as the lower bound for both $\lambda$ and $\theta$ was due to the limitation of our numerical method. That is, we chose the values of $\lambda$ and $\theta$ such that their ratios varied within $0.2$ and $1$ consistent with our criteria but not as far as the observational and theoretical threshold would allow it in the squeezed limit. 
However, our numerical solution did not indicate any divergence in the asymptotic behaviour of bispectrum. 
\par
\begin{figure}[ht] 
     \centering
     \begin{subfigure}[b]{0.5\textwidth}
         \centering
         \includegraphics[width=0.9\linewidth]{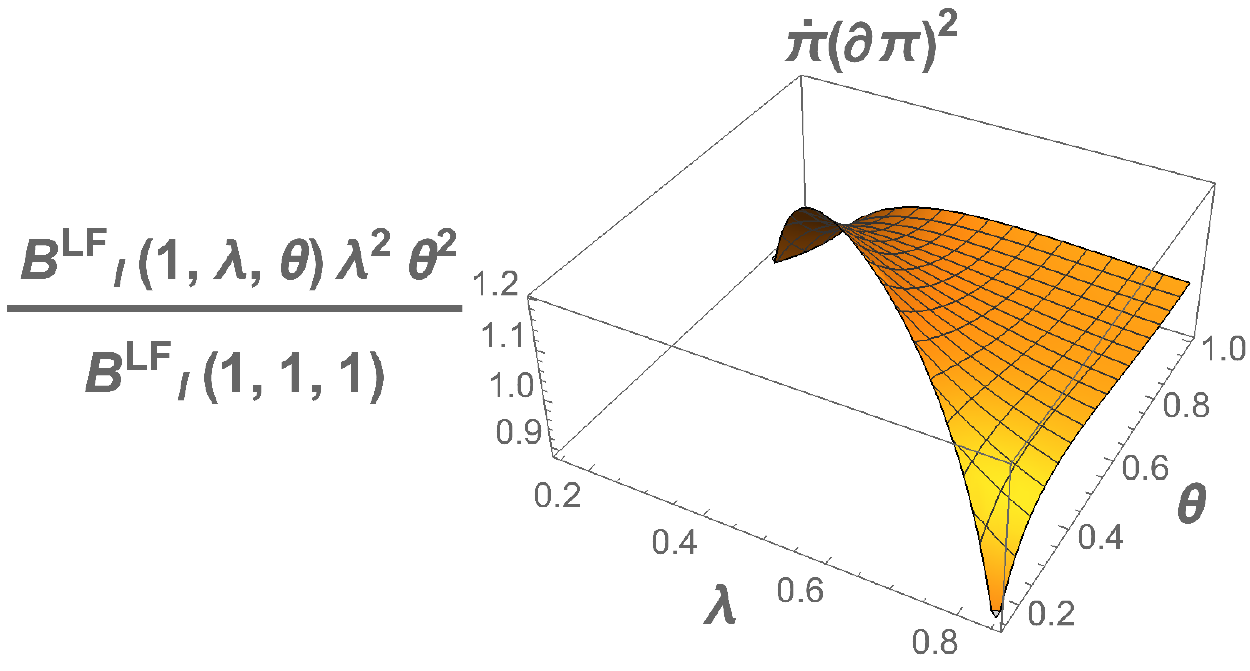}
     \end{subfigure}%
  \begin{subfigure}[b]{0.5\textwidth}
         \centering
         \includegraphics[width=0.9\linewidth]{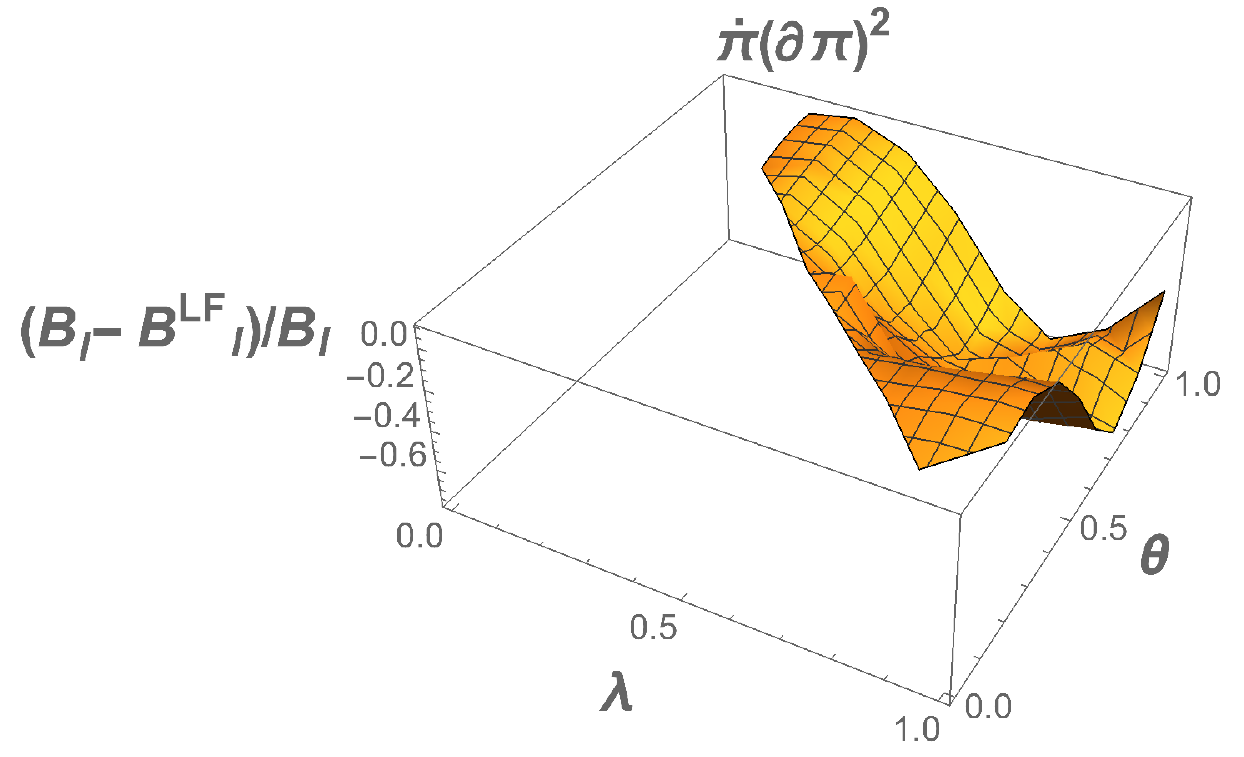}
     \end{subfigure}
     \caption{Left plot represents the shape of the best fit linear model for $\dot{\pi}(\partial\pi)^2$ to  $T^{{}^{\rm local}}$, $T^{{}^{\rm equil.}}$ and $T^{{}^{\rm ortho.}}$ with coefficients given in table \eqref{tb:fnldotpi} and right plot is the residual bispectrum after subtracting the best fit. The mode functions correspond to values of $\alpha_0=-0.18$ and $\beta_0=\alpha_0^2/4$.}
    \label{fig:bestfitdotBI}
\end{figure}

\begin{figure}[ht] 
     \centering
     \begin{subfigure}[b]{0.5\textwidth}
         \centering
         \includegraphics[width=0.9\linewidth]{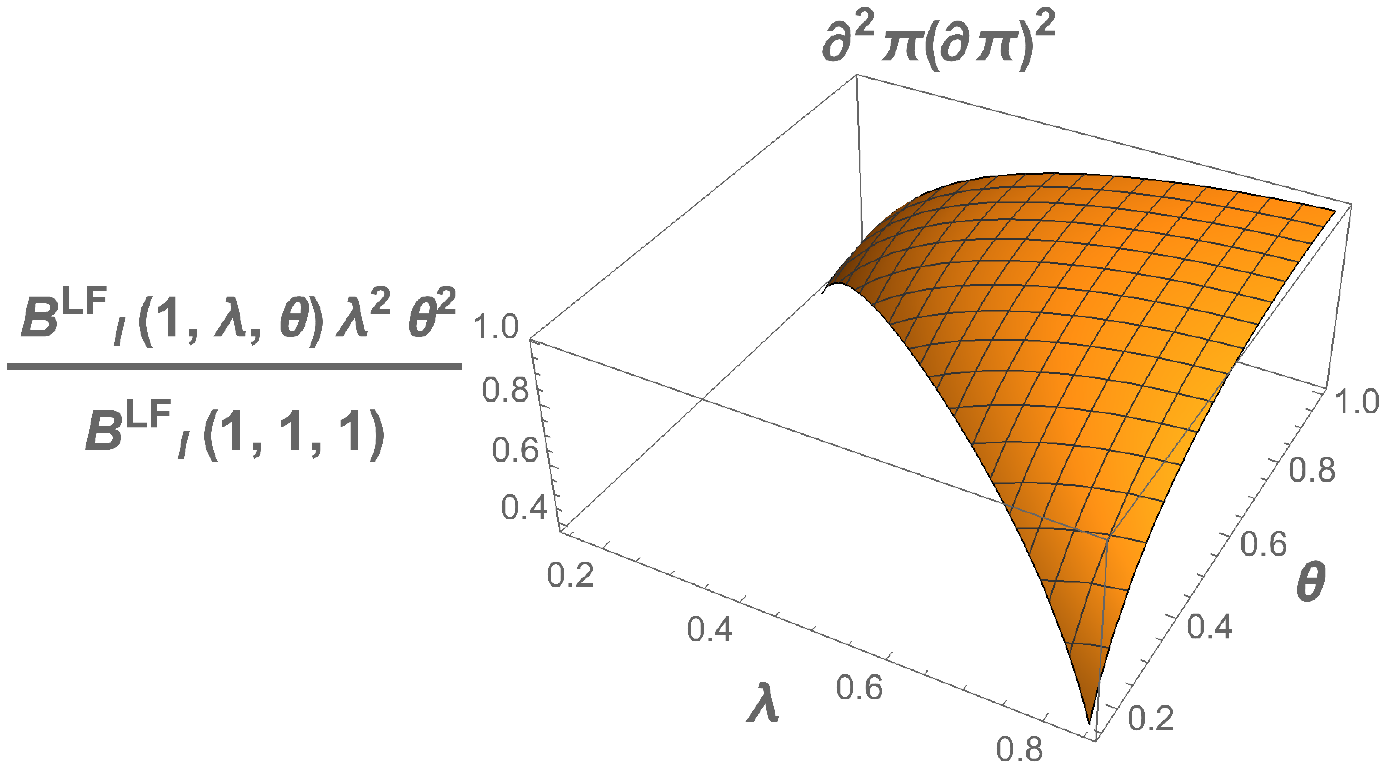}
     \end{subfigure}%
  \begin{subfigure}[b]{0.5\textwidth}
         \centering
         \includegraphics[width=0.9\linewidth]{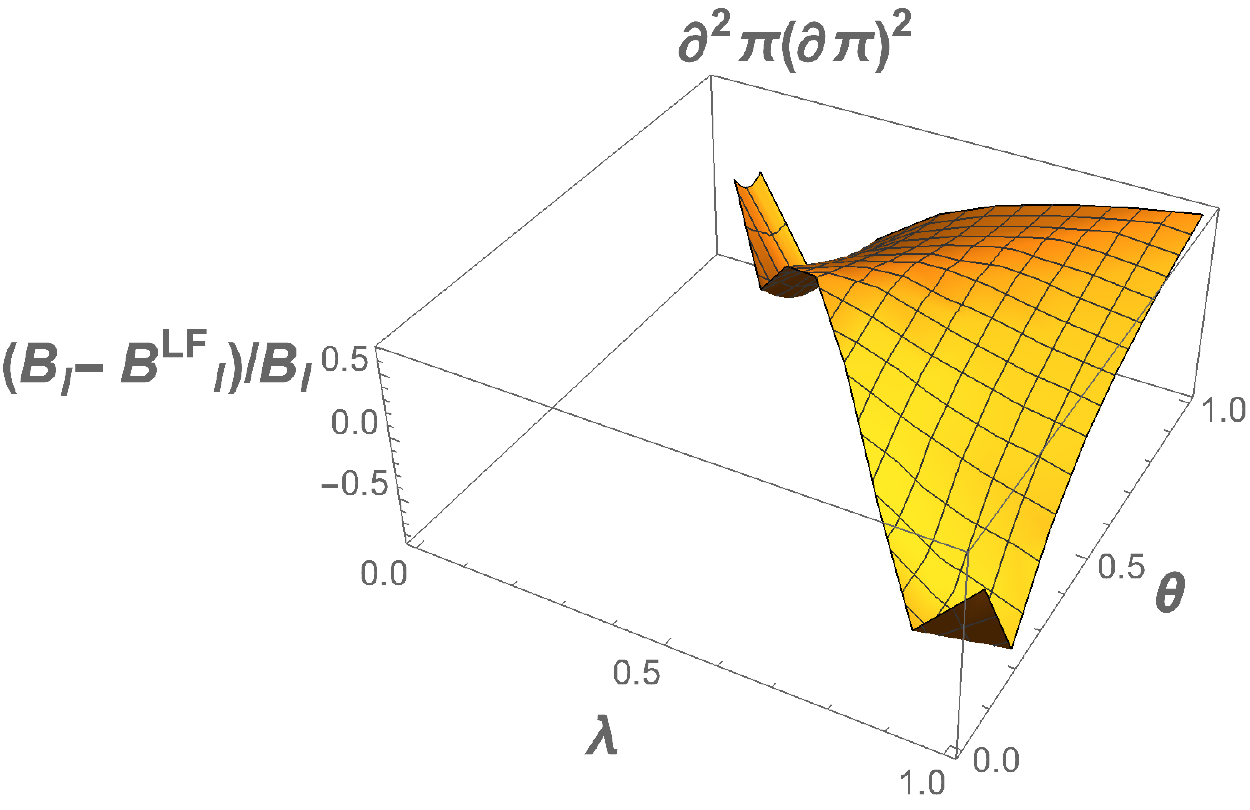}
     \end{subfigure}
     \caption{Left plot represents the shape of best fit linear model for $\partial^2\pi(\partial\pi)^2$ to  $T^{{}^{\rm local}}$, $T^{{}^{\rm equil.}}$ and $T^{{}^{\rm ortho.}}$ with coefficients given in table \eqref{tb:fnlpar2pi} and right plot is the residual bispectrum after subtracting the best fit. The mode functions correspond to values of $\alpha_0=-0.18$ and $\beta_0=\alpha_0^2/4$.}
    \label{fig:bestfitpar2BI}
\end{figure}

 Figure \eqref{fig:bestfitdotBI} represents the shape of the best fit linear model for $\dot{\pi}(\partial\pi)^2$ to $T^{{}^{\rm local}}$, $T^{{}^{equil.}}$ and $T^{{}^{ortho.}}$ with coefficients $\frac{f_{{}_{\rm NL}}^{{}^{(I)\dots}}\gamma_{{}_S}^2}{\tilde{q}_I}$ given in table \eqref{tb:fnldotpi} and the residual bispectrum after subtracting the best fit. The mode functions in this case correspond to the values of $\alpha_0=-0.18$ and $\beta_0=\alpha_0^2/4$. Similarly, figure \eqref{fig:bestfitpar2BI} displays the best linear fit model to the same templates for $\partial^2\pi(\partial\pi)^2$ operator and the residual bispectrum, corresponding to the same values of $\alpha_0$ and $\beta_0$ where obtained values of $\frac{f_{{}_{\rm NL}}^{{}^{(I)\dots}}\gamma_{{}_S}^2}{\tilde{q}_I~}$ are given in table \eqref{tb:fnlpar2pi}. 

\begin{table}[h] 
    \centering
\begin{tabularx}{0.7\textwidth} { 
  | >{\raggedright\arraybackslash}X 
  | >{\centering\arraybackslash}X 
  | >{\raggedleft\arraybackslash}X | }
 \hline
  \centering $\dot{\pi}(\partial\pi)^2$ & Estimate & Standard Error \\
 \hline
  \centering \tiny{$f_{{}_{\rm NL}}^{{}^{\rm (I)local}}\gamma_{{}_S}^2/\tilde{q}_I$} & -65  & 2.9\\
 \hline
  \centering \tiny{$f_{{}_{\rm NL}}^{{}^{\rm (I)equil.}}\gamma_{{}_S}^2/\tilde{q}_I$} &  881 & 16 \\\hline
   \centering \tiny{$f_{{}_{\rm NL}}^{{}^{\rm (I)ortho}}\gamma_{{}_S}^2/\tilde{q}_I$} &  -339 & 7  \\\hline
\end{tabularx}
\caption{The coefficients of the best fit linear model for $\dot{\pi}(\partial\pi)^2$  to $T^{{}^{\rm local}}$, $T^{{}^{\rm equil.}}$ and $T^{{}^{ortho.}}$ corresponding to values of $\alpha_0=-0.18$ and $\beta_0=\alpha_0^2/4$.}
\label{tb:fnldotpi}
\end{table}
 \begin{table}[ht]
    \centering
\begin{tabularx}{0.7\textwidth} { 
  | >{\raggedright\arraybackslash}X 
  | >{\centering\arraybackslash}X 
  | >{\raggedleft\arraybackslash}X | }
 \hline
  \centering \tiny{$\partial^2\pi(\partial\pi)^2$} & Estimate & Standard Error \\
 \hline
  \centering \tiny{$f_{{}_{\rm NL}}^{{}^{(I)\rm local}}\gamma_{{}_S}^2/\tilde{q}_I$} & -282  & 29\\
 \hline
  \centering \tiny{$f_{{}_{\rm NL}}^{{}^{(I)\rm equil.}}\gamma_{{}_S}^2/\tilde{q}_I$} &  3610 & 344  \\\hline
   \centering \tiny{$f_{{}_{\rm NL}}^{{}^{(I){\rm ortho.}}}\gamma_{{}_S}^2/\tilde{q}_I$} &  -1033 & 99  \\\hline
\end{tabularx}
\caption{The coefficients of the best fit linear model for $\partial^2\pi(\partial\pi)^2$  to $T^{{}^{\rm local}}$, $T^{{}^{\rm equil.}}$ and $T^{{}^{\rm ortho.}}$ corresponding to values of $\alpha_0=-0.18$ and $\beta_0=\alpha_0^2/4$.}
\label{tb:fnlpar2pi}
\end{table}
\begin{table}[b]
        \begin{tabularx}{0.9\textwidth}{|l|X|}
        \hline
        Constraints& Note\\\hline
        $\alpha_0 < 0$& Regions where the power spectrum is enhanced. Note that for $\alpha_0 > 0$ the power spectrum is suppressed (not interesting) \\\hline
        $\beta_0 > 0$& To avoid tachyonic instabilities (note $\delta_3 >0$)\\\hline
        $\Delta_{{}_{\zeta}} \sim 10^{-9}$& The constraint from power spectrum of the CMB observation ($1\lesssim \gamma_{{}_S} \lesssim 14$).\\\hline
        $H > \frac{\sqrt{A_1}}{M_{{}_{\rm Pl}}}$& Decoupling limit\\\hline
        $0\leq c_{{}_{T}}^2 \leq 1$& Speed of sound for the tensor modes (Note:  $\bar{M}_3^2 < 0$)\\\hline
        $0\leq c_{{}_{S}}^2 = \frac{F_1}{A_1}\leq 1$ &  To avoid superluminality in the regime $\omega^2\sim k^2$ \\\hline
        $A_1\geq 0 $& To avoid ghosts at the leading order\\\hline
        $\dot{H}<0$& Inflation satisfies the null energy condition\\\hline
        $\tilde{q}_i\leq 10^{-1}$& Constraints from $f_{{}_{\rm NL}}$ measurements\\\hline
        \end{tabularx}
        \caption{List of the constraints applied to the EEFToI models}
        \label{tb:constraints}
    \end{table}
\par We now try to explore the allowed regions of the parameters for EEFToI models or rather impose constraints on them, based on the observed constraints on the size of the non-Gaussianity. The most recent CMB observation by Planck collaboration \cite{planckcollaboration2019planck} the following bounds on the size of the non-Gaussianities for the local, equilateral and orthogonal templates:
$f^{{}^{\rm local}}_{{}_{\rm NL}}\simeq -0.9\pm 5.1$, $f^{{}^{\rm equil.}}_{{}_{\rm NL}}\simeq -26\pm 47$ and $f^{{}^{\rm ortho.}}_{{}_{\rm NL}}\simeq -38\pm 24$. Now to make the connection to our model, we make a comparison between these observational constraints for different shapes and the best linear fit values of $f_{{}_{\rm NL}}^{{}^{\rm (I)}}\gamma_{{}_S}^2/\tilde{q}_I$ to infer bounds on value of $\tilde{q}_I$. We then translate this into constraints on the parameters of our models in the action by substituting for different mass and coupling parameters in $\gamma_{{}_S}^2$ and $\tilde{q}_i$.
We require for each shape estimated $f_{{}_{\rm NL}}^{{}^{(I)}}$ be less than reported value, which leads to 
\begin{equation}
    \left|\frac{f_{{}_{\rm NL}}^{{}^(I)}\gamma_{{}_S}^2}{\tilde{q}_I}\right|^{{}^{\rm estimate}}< \frac{|f_{{}_{\rm NL}}^{{}^{\rm reported}}|\gamma_{{}_S}^2}{|\tilde{q}_I|} \implies |\tilde{q}_I| < \frac{|f_{{}_{\rm NL}}^{{}^{\rm reported}}|}{\left|\frac{f_{{}_{\rm NL}}^{(I)}\gamma_{{}_S}^2}{\tilde{q}_I}\right|^{{}^{\rm estimate}}}\gamma_{{}_S}^2 \lesssim 10^{-2}-10^{-3}\gamma_{{}_S}^2
\end{equation}
 \begin{figure}[t]
    \centering
    \begin{subfigure}[b]{1\textwidth}
        \centering
        \includegraphics[width=\textwidth]{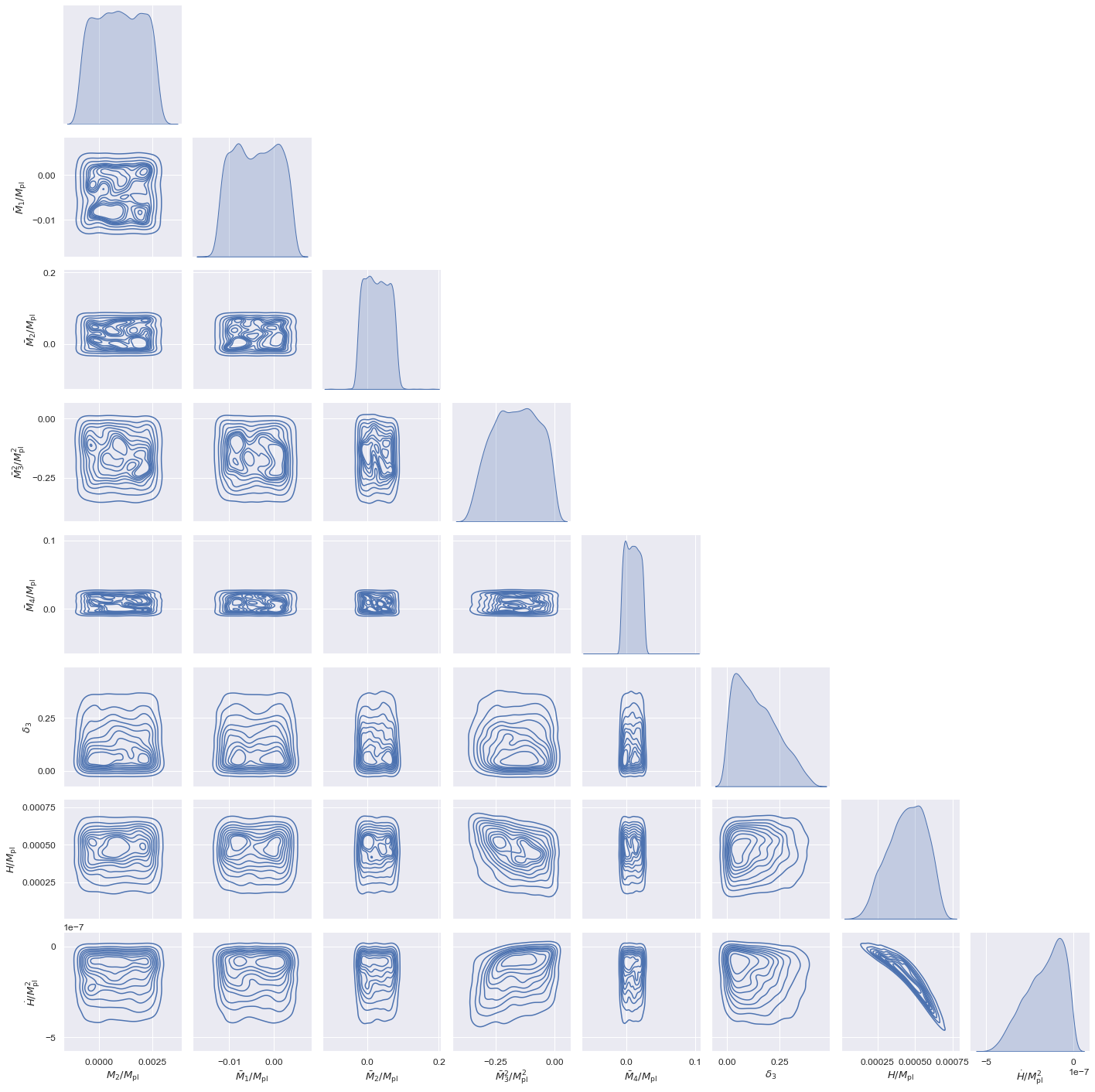}
    \end{subfigure}
    \caption{Parameters search of EEFToI model under the constraints listed in table \eqref{tb:constraints}. We searched for allowed values of $M_2, \bar{M}_1, \bar{M}_2, \bar{M}_3^2, \bar{M}_4, \delta_3, H, \dot{H}$. Furthermore, we set an additional cutoff at $M_{{}_{\rm Pl}}$, except for $\delta_3$ which is dimensionless. Although the points in the plots are scattered, this is due to the method  carried out for searching through parameters. Instead of searching the entire grid, which would take enormous computation time, we took a random search method. Note that the histogram does not represent the probability of each parameter. It just means that the algorithm searched in those respective regions more frequently.}
    \label{fig:parameterc1}
\end{figure}

Therefore, we can crudely set a bound on $\tilde{q}_I\lesssim~ 10^{-2}-10^{-3}\ \gamma_{{}_S}^2$. In this case substituting larger than unity values such as $\gamma_{{}_S} \sim 14$ results into a not very tight bound of $|\tilde{q}_I|\lesssim 10^{-1}$. 
Now to address what this constraint mean in terms of the original couplings and coefficients in EEFToI action, and whether there are corners of the parameter space that lead to scenarios we have explored, we run a search algorithm in the parameter space under the list of all the criteria we have set for the model. Table \eqref{tb:constraints} shows the list of all these  conditions and constraints. The primary parameters that were included in EEFToI were $M_2, \bar{M}_1, \bar{M}_2, \bar{M}_3, \bar{M}_4, \delta_3, H, M_{{}_{\rm Pl}}^2\dot{H}$.

Figure \eqref{fig:parameterc1} displays the result of a search algorithm we performed to find valid values of parameters of EEFToI imposing conditions \eqref{tb:constraints}. We also have imposed an additional cutoff on all of the mass dimension parameters to be below the Planck scale. That is to avoid break down of semi-classical gravity approximation. Figure \eqref{fig:parameterc1}, visualizes the result of the search algorithm indicating the values of parameters that were found to satisfy the above constraints. Therefore, we can infer that there are allowed regions in the mass and coupling parameter space, arising from the sixth order polynomial  dispersion relations \cite{Ashoorioon:2017toq} and leading to a sensible effective field theory without violating the current observations constraints on non-gaussianity.

\section{Conclusion}\label{sec:conc}
In earlier works, we had examined the power spectrum for some interesting corners of the parameter space in the EEFToI models with the dispersion relation $\omega^2 \propto k^6$ when the wavelength of the mode is very small \cite{Ashoorioon:2017toq,Ashoorioon:2018uey}. The dispersion relation then evolves into quartic and quadratic before the modes exit the horizon. In this work, we studied the bispectrum for these models in the regime where the amplification of the power spectrum due to the modified dispersion relation is more significant. One implications of taking into account such possibilities in EEFToI is that they can lead to lower values of $H$ during inflation \eqref{eq:powerspectrum2} and still be consistent with the observed value of amplitude for scalar perturbations. This widens the possibilities of viable low energy inflationary models. 

Our power counting arguments in section \eqref{section:ng} indicated that the most pronounced and possibly dangerous contribution of the cubic interaction terms are coming from $\dot\pi(\partial_i\pi)^2$ and $\partial_i^2\pi(\partial_j\pi)^2$, we computed their impact. 
We found the triangle shapes corresponding to peaks and cresses in the bispectrum (Figure \ref{fig:BI}) are quite distinct from other inflationary models. In particular, we made a linear fit of the shapes of the bispectra for these two operators, to the three known and widely used templates of local, equilateral and orthogonal shapes. Interestingly, the major contributions to the bispectra for these two cubic interaction operators was projected on the equilateral configuration, whereas in the other EFT of inflation with non-Bunch-Davies vacuum, it is often the orthogonal template which is dominant. 
This calculation also allowed us to inspect the strong coupling constraints and the restrictions due to the size of non-Gaussianity from the CMB observations. We translated the reported observational constraints from Planck Collaboration \cite{planckcollaboration2019planck} in values of $f_{{}_{\rm NL}}$ in terms of the parameters of the model. This lead to an order of magnitude estimation for the bounds on variables $\tilde{q}_I$ which include the masses and couplings in the model defined in \eqref{defq} as $\tilde{q}_I\lesssim 10^{-1}$. Finally, we used a search algorithm and found that there are allowed regions of parameter space where EEFToI description with $\omega^2 \propto k^6$ is sensible and interesting. For future, it is intriguing to perform a direct comparison of our numerical template for bispectrum with the data itself and investigate its signatures.  Another aspect that we leave for future work is analysing the trispectrum and specifically the impact and signatures of the term $(\partial_i\pi)^4$ in the Extended EFT of inflation.  
 Furthermore, we like to add that the terms we have included in this analysis are not the exhaustive list of all the possible terms consistent with spatial diffeomorphisms in unitary gauge upto mass dimension 4. Our analysis has been more focused on terms leading to interesting $\omega^2 \propto k^6$ regimes and enhanced signatures of purely scalar bispectrum for such regimes. However, there are other possible regimes and also contractions of $g^{00}+1,\delta K_{\mu\nu}, \delta R_{\mu\nu\rho\sigma}, \nabla_\mu$ that through their quadratic or interaction contributing terms in the action, could potentially lead to interesting and distinctive observational signatures. 

\section{Acknowledgment}
A.A. has received funding/support from the European Union’s Horizon 2020 research and innovation programme under the Marie Skłodowska -Curie grant agreement No 860881-HIDDeN. G.G. and H.J.K. research is supported by the Discovery
Grant from Natural Science and Engineering Research Council of Canada (NSERC) and G.G.
is supported partly by Perimeter Institute (PI) as well. Research at PI is supported by the Government of
Canada through the Department of Innovation, Science and Economic Development Canada
and by the Province of Ontario through the Ministry of Research, Innovation and Science.

\appendix
\section{The Interaction Lagrangian for the EEFToI}
\label{AppendixA}
In the section we summarize all the cubic terms in the action $\int d^4x \sqrt{g}\sum_{d=3}^4\mathcal{L}_{2, d}$ generated through the new EEFToI terms \eqref{eq:actiontad} after setting $\delta_1 = \delta_2 = 0$.
\par The contributions from the operator proportional to $\bar{M}_4$ to the cubic interactions in the action, taking into account $\sqrt{g}$, goes as: 
\begin{dmath}
   a^3\nabla^\mu \delta g^{00}\nabla^\nu\delta K_{\mu\nu} \longrightarrow -\frac{3 H (\partial_i\pi)^2 \partial _j^2\dot\pi}{a}+\frac{H^2 (\partial_i\pi{})^2 \partial _j^2\pi}{a}+18 a H^3 \dot{\pi} (\partial_i\pi{})^2-14 a H^2 \dot{\pi} \partial_i\pi \partial _i\dot\pi-8 a H \dot{\pi} \ddot{\pi} \partial _i^2\pi+18 a^3 H^2 \dot\pi^2 \ddot{\pi}+\frac{\partial _j\pi \partial _i\partial _j\pi \partial _i\partial _k^2\pi}{a^3}+\frac{\partial_i\pi \partial _i\partial _j\pi \partial _k^2\partial _j\pi}{a^3}+\frac{2 \ddot{\pi} \partial _i\partial _j\pi{}^2}{a}+\frac{\partial _j\pi \partial _i\dot\pi  \partial _i\partial _j\dot\pi}{a}+\frac{5 \partial _i\dot\pi \partial _i\partial _j\pi \partial _j\dot\pi}{a}+\frac{\partial_i\pi \partial _i\dot\pi \partial _j\partial _j\dot\pi}{a}
\end{dmath}
\par The contributions from the operator proportional to the $\delta_3$ to the cubic interaction terms in the  action goes as:
\begin{dmath*}
   {a^3} \nabla^\mu\delta K_{\nu\mu}\nabla^\nu\delta {K^\sigma}_\sigma\longrightarrow 3 a \dot{\pi}  \left(\partial _i\pi \right){}^2 H^4+18 a^3 \dot{\pi}^2 \ddot{\pi}H^3-21 a \dot{\pi}  \partial _i \pi \partial _i\dot{\pi} H^3
   +6 a \dot{\pi}^2 \partial _i^2\pi  H^3-\frac{\left(\partial _i\pi \right){}^2 \partial_j^2\pi  H^3}{a}+3 a \dot{\pi}  \partial _i \pi \partial _i\ddot{\pi} H^2+\frac{4 \dot{\pi}  \partial _i \pi \partial _i\pi\partial _j^2\pi H^2}{a}
   -6 a \dot{\pi}  \ddot{\pi}  \partial _i^2\pi  H^2-3 a \dot{\pi}^2
   \partial _i^2\dot{\pi} H^2+\frac{2 \partial _i\dot{\pi} \partial
   _i\partial _j\pi \partial _j\pi  H^2}{a}+\frac{9 \partial _i^2\pi 
   \partial _j\pi  \partial _j\dot{\pi} H^2}{a}-\frac{4 \dot{\pi}  \partial _i^2\pi 
   \partial _j^2\pi  H^2}{a}+18 a^3 \dot{\pi}^3 \dot{H} H^2-\frac{3}{2} a
   \dot{\pi}  \left(\partial _i\pi \right){}^2 \dot{H} H^2+\frac{3 \ddot{\pi}  \left(\partial
   _i\partial _j\pi\right){}^2 H}{a}+\frac{3 \partial _i^2\pi 
   \left(\partial _j\dot{\pi}\right){}^2 H}{2 a}+
   \frac{5 \dot{\pi}  \partial _i\dot{\pi} \partial _i\partial _j^2\pi H}{a}+\frac{5 \partial _i\dot{\pi} \partial _i\left(\partial _j\dot{\pi}\right) \partial _j\pi  H}{2
   a}+\frac{17 \partial _i\dot{\pi} \partial _i\left(\partial _j\pi \right) \partial
   _j\dot{\pi} H}{2 a}+\frac{3 \partial _i^2\dot{\pi} \partial _j\pi 
   \partial _j\dot{\pi} H}{2 a}-\frac{4 \partial _i^2\pi  \partial _j\pi 
   \partial _j\partial _k^2\pi  H}{a^3}-\frac{\partial _i \pi \partial
   _i\ddot{\pi}  \partial _j^2\pi  H}{a}-\frac{\ddot{\pi}  \partial _i^2\pi  \partial
   _j^2\pi  H}{a}+\frac{2 \dot{\pi}  \partial _i^2\pi  \partial _j^2\dot{\pi}
   H}{a}+\frac{4 \partial _i^2\pi \left( \partial _j\partial _k\pi \right)^2\pi H}{a^3}+\frac{2 \left(\partial _i\partial _j\pi
   \right){}^2 \partial _k^2\pi  H}{a^3}-\frac{2 \partial _i \pi \partial
   _i\partial _j^2\pi \partial _k^2\pi  H}{a^3}
 -27 a \dot{\pi}^2 \partial _i^2\pi  \dot{H} H+\frac{3 \left(\partial _i\pi \right){}^2 \partial
   _j^2\pi  \dot{H} H}{2 a}
 \\-\frac{4 \dot{\pi}  \left(\partial _i\partial _j^2\pi \right)^2
  }{a^3}-\frac{7 \partial _i\partial _j\pi
    \partial _i\partial _k^2\pi \partial _j\dot{\pi}}{2
   a^3}-\frac{3 \partial _i \pi \partial _i\partial _k\dot{\pi}
   \partial _j^2\partial _k\pi}{2 a^3}-\frac{\partial _i^2\dot{\pi}
   \partial _j^2\partial _k\pi  \partial _k\pi }{a^3}-\frac{2 \partial
   _i^2\dot{\pi} \partial _j\partial _k\pi \partial _k\partial
   _j\pi }{a^3}-\frac{3 \partial _i^2\partial _j\pi \partial
   _j\dot{\pi} \partial _k^2\pi }{2 a^3}-\frac{\left(\partial _i\partial
   _j\pi \right){}^2 \partial _k^2\dot{\pi}}{a^3}-\frac{\partial
   _i^2\partial _j\pi \partial _j\pi  \partial _k^2\dot{\pi}}{2
   a^3}-\frac{2 \dot{\pi} \left(\partial _i^2\partial _j\pi \right)^2 }{a^3}+\frac{15 \dot{\pi}
   (\partial _j\partial _i\pi)^2 \dot{H}}{a}
\end{dmath*}
\par The contribution of the operator proportional to the $\delta_4$  in Lagrangian to the cubic interactions is given by:
\begin{dmath}
    {a^3}\nabla_\mu\delta {K^\mu}_\nu\nabla_\gamma\delta K^{\gamma\nu} \longrightarrow \frac{2 H \partial _j^2\pi  \left(\partial _i\partial _k\pi\right){}^2}{a^3}-\frac{2 H \partial _i \pi \partial _j^2\pi  \partial
   _i\partial _k^2\pi}{a^3}-\frac{6 \dot{\pi}  H^2 \left(\partial
   _i\partial _j\pi\right){}^2}{a}+\frac{4 H^2 \partial _i \pi \partial
   _j^2\pi  \partial _i\dot{\pi}}{a}-\frac{2 \dot{\pi}  H^2 \partial _i^2\pi 
   \partial _j^2\pi }{a}+\frac{4 \dot{\pi}  H \partial _i\partial _j^2\pi
   \partial _i\dot{\pi}}{a}+\frac{H \partial _j\pi  \partial _i\dot{\pi}
   \partial _i\partial _j\dot{\pi}}{a}+\frac{5 H \partial
   _i\partial _j\pi \partial _i\dot{\pi} \partial _j\dot{\pi}}{a}+\frac{H \partial _j^2\pi  \left(\partial _i\dot{\pi}\right){}^2}{a}+\frac{H \partial _i \pi \partial _i\dot{\pi} \partial
   _j^2\dot{\pi}}{a}-\frac{H^3 \left(\partial _i\pi \right){}^2 \partial _j^2\pi
   }{a}+3 \dot{\pi}  a H^4 \left(\partial _i\pi \right){}^2-6 \dot{\pi}  a H^3 \partial
   _i\pi  \partial _i\dot{\pi}+12 \dot{\pi}^2 a H^3 \partial _i^2\pi
   -2 \dot{\pi} a H^2 \left(\partial _i\dot{\pi}\right){}^2-\frac{2 \dot{\pi}  (\partial
   _i\partial _j^2\pi )^2
   }{a^3}-\frac{\partial _j\pi  \partial _i\partial _k^2\pi \partial
   _i\partial _j\dot{\pi}}{a^3}-\frac{5 \partial _i\partial _j\pi  \partial _i\partial _k^2\pi \partial _j\dot{\pi}}{a^3}-\frac{\partial _j^2\pi  \partial _i\partial _k^2\pi 
   \partial _i\dot{\pi}}{a^3}-\frac{\partial _j\pi  \partial _i^2\partial
   _j\pi  \partial _k^2\dot{\pi}}{a^3}+18 \dot{\pi}^3 a^3 H^4
\end{dmath}

\section{More Details about the Numerical Calculations of Bispectrum and the Variable Transformations} \label{numericalcalculations}
To calculate different contributions to $\left<\zeta(\textbf{k}_1)\zeta(\textbf{k}_2)\zeta(\textbf{k}_3)\right>$, we first substituted $\zeta$ for $\pi$ using $\pi \rightarrow \zeta = -H\pi$ and then transformed from cosmic time to conformal time $dt\rightarrow a\; d\tau$. For the $\dot\pi(\partial\pi)^2$ term, this results in factoring factors of $H$ and $a$ into the coefficients in the following way, 
\begin{equation}
   C_{\dot{\pi} (\partial\pi)^2}\dot{\pi} (\partial\pi)^2 \xrightarrow{}-\frac{1}{H^3}C_{\dot{\pi} (\partial\pi)^2}\dot\zeta(\partial\zeta)^2 \xrightarrow{} -\frac{1}{aH^3}C_{\dot{\pi} (\partial\pi)^2}\zeta'(\partial\zeta)^2.
\end{equation}
Therefore, we took track of these factors by defining
\begin{equation}
    C_{\zeta'(\partial\zeta)^2} = -\frac{1}{aH^3}C_{\dot{\pi} (\partial\pi)^2}=-\frac{1}{H^3}[-\delta_3H^4 +9\bar{M}_4H^3-2M_2^4]. 
\end{equation}
After evaluating all the commutation relations between the operators in the interaction terms and the external vertices, we found the contribution to the bispectrum to be
\begin{align}
   \left<\zeta(\textbf{k}_1)\zeta(\textbf{k}_2)\zeta(\textbf{k}_3)\right>_{\dot\pi(\partial\pi)^2}=-i\int_{-\infty}^{\tau_f} a(\tau')d\tau' C_{\zeta'(\partial\zeta)^2}&\;2(2\pi)^3\delta^3(\textbf{k}_3+\textbf{k}_2+\textbf{k}_1)\times\nonumber\\f_{k_1}(\tau_f){f_{k_1}^*}'(\tau')f_{k_2}(\tau_f){f_{k_2}^*}(\tau')&f_{k_3}(\tau_f){f_{k_3}^*}(\tau'){\textbf{k}}_2\cdot{\textbf{k}}_3+ \text{\text{cycl.}} + \text{c.c.},
\end{align} where $f$ as we had defined in \eqref{mfunction} represents the mode function for $\zeta(\textbf{k})$. Next, we substitute for $f$ and $f'$ in terms of the canonical variable $u_k(\tau)$ which satisfies the Wronskian condition and had been computed numerically from quadratic action. The substitution relations are given by
\begin{align*}
\sqrt{A_1}f &= -\frac{H}{a}u\\
\sqrt{A_1}f'&= -\frac{H}{a}\frac{du}{d\tau} +H^2u
\end{align*} 
to obtain
\begin{align}
   \left<\zeta(\textbf{k}_1)\zeta(\textbf{k}_2)\zeta(\textbf{k}_3)\right>_{\dot\pi(\partial\pi)^2}&=&-i\int_{-\infty}^{\tau_f}a(\tau')d\tau'\left(\frac{H^6}{a(\tau_f)^3a(\tau')^3 A_1^3}\right)\;C_{\zeta'(\partial\zeta)^2}\;2(2\pi)^3\delta^3(\textbf{k}_3+\textbf{k}_2+\textbf{k}_1)\nonumber\\
   &&u_{k_1}(\tau_f){u_{k_1}^*}'(\tau')u_{k_2}(\tau_f){u_{k_2}^*}(\tau')u_{k_3}(\tau_f){u_{k_3}^*}(\tau'){\textbf{k}}_2\cdot{\textbf{k}}_3+ \text{cycl.} + \text{c.c.}\nonumber\\
    &&+i\int_{-\infty}^{\tau_f}a(\tau')d\tau'\left(\frac{H^{7}}{a(\tau_f)^3a(\tau')^2 A_1^3}\right)C_{\zeta'(\partial\zeta)^2}\;2(2\pi)^3\delta^3(\textbf{k}_3+\textbf{k}_2+\textbf{k}_1)\nonumber\\
   &&u_{k_1}(\tau_f){u_{k_1}^*}(\tau')u_{k_2}(\tau_f){u_{k_2}^*}(\tau')u_{k_3}(\tau_f){u_{k_3}^*}(\tau'){\textbf{k}}_2\cdot{\textbf{k}}_3+ \text{cycl.} + \text{c.c.}
\end{align} 
Next, changing the integration variable to the dimensionless variable $x_i=c_sk_i\tau$, and substituting for $a=-1/H\tau= -c_sk_i/Hx_i$ and defining $\tilde{u}(x)= \sqrt{c_s k}\,u_k(\tau)$\footnote{This is due to the symmetry of our ODE and normalization under this change of variable. } as well as $x_f=c_sk_1\tau_f$, we arrive at. 
\begin{align}
   \left<\zeta(\textbf{k}_1)\zeta(\textbf{k}_2)\zeta(\textbf{k}_3)\right>_{\dot\pi(\partial\pi)^2}=&i2(2\pi)^3\delta^3(\textbf{k}_3+\textbf{k}_2+\textbf{k}_1)\frac{1}{\sqrt{c_s^6k_1^2k_2^2k_3^2}}\frac{H^{11}x_f^3}{(c_sk_1)^5 A_1^3} \left [\int_{-\infty}^{x_{f}}(x_1')^2 dx_1'\; \right. \nonumber\\& 
  C_{\zeta'(\partial\zeta)^2}\; \tilde{u}(x_f){\tilde{u}^*}_{,x}(x_1')\tilde{u}(\frac{k_2}{k_1}x_f){\tilde{u}^*}(\frac{k_2}{k_1}x_3')\tilde{u}(\frac{k_3}{k_1}x_f){\tilde{u}^*}(\frac{k_2}{k_1} x_1') \nonumber\\&{\textbf{k}}_2\cdot{\textbf{k}}_3
  +\int_{-\infty}^{x_f}(x_1') dx_1'\;C_{\zeta'(\partial\zeta)^2} \tilde{u}(x_f){\tilde{u}^*}(x_1')\tilde{u}(\frac{k_2}{k_1}x_f){\tilde{u}^*}(\frac{k_2}{k_1}x_1') \nonumber\\
   & 
  \tilde{u}(\frac{k_3}{k_1}x_f){\tilde{u}^*}(\frac{k_3}{k_1}x_1'){\textbf{k}}_2\cdot{\textbf{k}}_3\bigg]+ \text{cycl.} + \text{c.c.}
\end{align}
Applying the Cosine Law, 
\begin{equation}
    k_1^2 = k_2^2+k_3^2 - 2 k_3k_2\cos(\vartheta) \Rightarrow \textbf{k}_2\cdot\textbf{k}_3 = k_3k_2\cos(\vartheta) = -\frac{k_1^2 - k_3^2 - k_2^2}{2},
\end{equation}
and substituting for $C_{\zeta'(\partial\zeta)^2}$ lead to 
\begin{align}
   \left<\zeta(\textbf{k}_1)\zeta(\textbf{k}_2)\zeta(\textbf{k}_3)\right>_{\dot\pi(\partial\pi)^2}=&i(2\pi)^3\delta^3(\textbf{k}_3+\textbf{k}_2+\textbf{k}_1)\frac{(k_1^2 - k_3^2 - k_2^2)}{\sqrt{c_s^6k_1^2k_2^2k_3^2}}\frac{H^{8}x_f^3}{(c_sk_1)^5 A_1^3} \nonumber\\& [-\delta_3H^4 +9\bar{M}_4H^3-2M_2^4] \left [\int_{-\infty}^{x_{f}}(x_1')^2 dx_1'\; \right.  
  \tilde{u}(x_f){\tilde{u}^*}_{,x}(x_1')\nonumber\\&\tilde{u}(\frac{k_2}{k_1}x_f){\tilde{u}^*}(\frac{k_2}{k_1}x_3')\tilde{u}(\frac{k_3}{k_1}x_f){\tilde{u}^*}(\frac{k_3}{k_1}x_1') \nonumber\\&
  +\int_{-\infty}^{x_f} x_1' dx_1'\;\tilde{u}(x_f){\tilde{u}^*}(x_1') \tilde{u}(\frac{k_2}{k_1}x_f){\tilde{u}^*}(\frac{k_2}{k_1}x_1') \nonumber\\
   & 
  \tilde{u}(\frac{k_3}{k_1}x_f){\tilde{u}^*}(\frac{k_3}{k_1}x_1')\bigg]+ \text{cycl.} + \text{c.c.}
\end{align}
Now by defining parameters $\lambda \equiv \frac{k_2}{k_1}$ and $\theta \equiv \frac{k_3}{k_1}$ which determine the shape of the triangles formed by $\textbf{k}_1,~ \textbf{k}_2 $ and $\textbf{k}_3$ we can rewrite the above expression as 
\begin{align}
   \left<\zeta(\textbf{k}_1)\zeta(\textbf{k}_2)\zeta(\textbf{k}_3)\right>_{\dot\pi(\partial\pi)^2}=&(2\pi)^3\delta^3(\textbf{k}_3+\textbf{k}_2+\textbf{k}_1)\frac{H^{8}}{c_s^8 A_1^3} [-\delta_3H^4 +9\bar{M}_4H^3-2M_2^4]  \nonumber\\& \left [i\frac{(1 - \lambda^2 - \theta^2)x_f^3}{k_1^6\lambda\theta}\left\{\int_{-\infty}^{x_{f}}(x_1')^2 dx_1'\; \right. \right. 
  \tilde{u}(x_f){\tilde{u}^*}_{,x}(x_1')\nonumber\\&\tilde{u}(\lambda \;x_f){\tilde{u}^*}(\lambda \; x_1')\tilde{u}(\theta\; x_f){\tilde{u}^*}(\theta\; x_1') \nonumber\\&
  +\int_{-\infty}^{x_f}(x_1') dx_1'\; \tilde{u}( x_f){\tilde{u}^*}(x_1')\tilde{u}(\lambda \;x_f){\tilde{u}^*}(\lambda \;x_1') \nonumber\\
   & 
  \tilde{u}(\theta\; x_f){\tilde{u}^*}(\theta\; x_1')+ \text{c.c.}\bigg\}\bigg]+ \text{cycl.} 
\end{align}
Similar procedure was performed for evaluating the contributions from the $\partial^2\pi(\partial\pi)^2$ term.

\section{Calculations of Bispectrum for the Local-type Non-Gaussianity}\label{define FNL}
For the local-type NG, we characterise the size of the non-Gaussianty by parameter $f_{{}_{\rm NL}}$ as
\begin{equation}\label{localng}
    \zeta(x) = \zeta_{{}_{\rm L}}(x) +\frac{3}{5}f_{{}_{\rm NL}}\zeta_{{}_{\rm L}}^2(x),
\end{equation}
with
\begin{equation}
    \left<\zeta_{{}_{\rm L}}(x)\right> = 0.
\end{equation}
The Fourier transform of $\zeta_{{}_{\rm L}}$ is then obtained by 
\begin{eqnarray}
    \zeta(\textbf{k}) &=& \int d^3x\; e^{-ikx}\zeta(x) = \int d^3x\; e^{-ikx} \left(\zeta_{{}_{\rm L}}(x) + \frac{3}{5}f_{{}_{\rm NL}}\zeta_{{}_{\rm L}}^2(x)\right)\nonumber \\ &
 =&\int d^3x\; e^{-ikx}\left(\frac{1}{(2\pi)^{3}}\int d^3\tilde{k}\; e^{i\tilde{k}x}\zeta_{{}_{\rm L}}(\tilde{\textbf{k}}) + \frac{3}{5}f_{{}_{\rm NL}}\frac{1}{(2\pi)^6}\int d^3\tilde{k}\; e^{i\tilde{k}x}\zeta_{{}_{\rm L}}(\tilde{\textbf{k}})\int d^3k'\; e^{i{k'}x}\zeta_{{}_{\rm L}}(\textbf{k}')\right)
    \nonumber \\ & =&\left(\int d^3\tilde{k}\;\delta^3(\textbf{k}-\tilde{\textbf{k}}) \zeta_{{}_{\rm L}}(\tilde{\textbf{k}}) +\frac{3}{5}f_{{}_{\rm NL}}\frac{1}{(2\pi)^{3}}\int d^3\tilde{k}d^3k'\; \delta^3(\textbf{k}-\textbf{k}'-\tilde{\textbf{k}})\zeta_{{}_{\rm L}}(\tilde{\textbf{k}}) \; \zeta_{{}_{\rm L}}(\textbf{k}')\right)
\nonumber \\ & =& \zeta_{{}_{\rm L}}(\textbf{k}) + \frac{3}{5}f_{{}_{\rm NL}}\frac{1}{(2\pi)^{3}}\int d^3k'\; \zeta_{{}_{\rm L}}(\textbf{k}-\textbf{k}') \; \zeta_{{}_{\rm L}}(\textbf{k}') \end{eqnarray}
Substituting this for $\zeta(\textbf{k})$ in three-point function leads to 
\begin{align}
    \left<\zeta(\textbf{k}_1)\zeta(\textbf{k}_2)\zeta(\textbf{k}_3)\right> =& \left<\zeta_{{}_{\rm L}}(\textbf{k}_1)\zeta_{{}_{\rm L}}(\textbf{k}_2)\zeta_{{}_{\rm L}}(\textbf{k}_3)\right>\nonumber \\&+\frac{3}{5}f_{{}_{\rm NL}}\frac{1}{(2\pi)^3}\left<\zeta_{{}_{\rm L}}(\textbf{k}_1)\zeta_{{}_{\rm L}}(\textbf{k}_2)\int d^3k'\; \zeta_{{}_{\rm L}}(\textbf{k}_3-\textbf{k}') \; \zeta_{{}_{\rm L}}(\textbf{k}')\right>+\dots
\end{align}
Using Isserlis' theorem, the first term becomes
\begin{equation}
    \left<\zeta_{{}_{\rm L}}(\textbf{k}_1)\zeta_{{}_{\rm L}}(\textbf{k}_2)\zeta_{{}_{\rm L}}(\textbf{k}_3)\right>=\left<\zeta_{{}_{\rm L}}(\textbf{k}_1)\right>\left<\zeta_{{}_{\rm L}}(\textbf{k}_2)\zeta_{{}_{\rm L}}(\textbf{k}_3)\right>=\dots=0.
\end{equation}
The second term can also be calculated using Wick's theorem
\begin{align}
    \left<\zeta_{{}_{\rm L}}(\textbf{k}_1)\zeta_{{}_{\rm L}}(\textbf{k}_2)\int d^3k'\; \zeta_{{}_{\rm L}}(\textbf{k}_3-\textbf{k}') \; \zeta_{{}_{\rm L}}(\textbf{k}')\right>=&\int d^3k'\;\left<\zeta_{{}_{\rm L}}(\textbf{k}_1)\zeta_{{}_{\rm L}}(\textbf{k}_2) \zeta_{{}_{\rm L}}(\textbf{k}_3-\textbf{k}') \; \zeta_{{}_{\rm L}}(\textbf{k}')\right>\nonumber\\
    =&\int d^3k'\;\left<\zeta_{{}_{\rm L}}(\textbf{k}_1)\zeta_{{}_{\rm L}}(\textbf{k}_2)\right>\left< \zeta_{{}_{\rm L}}(\textbf{k}_3-\textbf{k}') \; \zeta_{{}_{\rm L}}(\textbf{k}')\right>
    \nonumber\\&+\int d^3k'\;\left<\zeta_{{}_{\rm L}}(\textbf{k}_1)\zeta_{{}_{\rm L}}(\textbf{k}')\right>\left<  \zeta_{{}_{\rm L}}(\textbf{k}_2)\zeta_{{}_{\rm L}}(\textbf{k}_3-\textbf{k}') \right>
   \nonumber\\& +\int d^3k'\;\left<\zeta_{{}_{\rm L}}(\textbf{k}_2)\zeta_{{}_{\rm L}}(\textbf{k}')\right>\left<  \zeta_{{}_{\rm L}}(\textbf{k}_1)\zeta_{{}_{\rm L}}(\textbf{k}_3-\textbf{k}') \right>
\end{align}
Next, taking into account that 
\begin{equation}
    \left<\zeta_{{}_{\rm L}}(\textbf{k}_1)\zeta_{{}_{\rm L}}(\textbf{k}_2)\right> = (2\pi)^3\delta^3(\textbf{k}_1+\textbf{k}_2)P_{{}_{\zeta}}(k_1)\end{equation}
and 
\begin{equation}\left<\zeta_{{}_{\rm L}}(\textbf{k}_3-\textbf{k}') \; \zeta_{{}_{\rm L}}(\textbf{k}')\right> = (2\pi)^3\delta^3(\textbf{k}_3)P_{{}_\zeta}(k').\end{equation}
the first term reduced to 
\begin{equation}
    \int d^3k'\;\left<\zeta_{{}_{\rm L}}(\textbf{k}_1)\zeta_{{}_{\rm L}}(\textbf{k}_2)\right>\left< \zeta_{{}_{\rm L}}(\textbf{k}_3-\textbf{k}') \; \zeta_{{}_{\rm L}}(\textbf{k}')\right> = \int d^3k'\;(2\pi)^6\delta^3(\textbf{k}_1+\textbf{k}_2)\delta^3(\textbf{k}_3)P_{{}_\zeta}(k_1)P_{{}_\zeta}(k')
\end{equation}
which does not contribute to three point function for $k_3\neq 0 $.
The second and the third terms are the same as
\begin{align*}
    \int d^3k'\;\left<\zeta_{{}_{\rm L}}(\textbf{k}_1)\zeta_{{}_{\rm L}}(\textbf{k}')\right>\left<  \zeta_{{}_{\rm L}}(\textbf{k}_2)\zeta_{{}_{\rm L}}(\textbf{k}_3-\textbf{k}') \right> =& \int d^3k'\;(2\pi)^6\delta^3(\textbf{k}_1+\textbf{k}')P(k_1)\delta^3(\textbf{k}_2+\textbf{k}_3-\textbf{k}')P_{{}_\zeta}(k_2)\\
    =&(2\pi)^6\delta^3(\textbf{k}_2+\textbf{k}_3+\textbf{k}_1)P_{{}_\zeta}(k_1)P_{{}_\zeta}(k_2),\\
    \int d^3k'\;\left<\zeta_{{}_{\rm L}}(\textbf{k}_2)\zeta_{{}_{\rm L}}(\textbf{k}')\right>\left<  \zeta_{{}_{\rm L}}(\textbf{k}_1)\zeta_{{}_{\rm L}}(\textbf{k}_3-\textbf{k}') \right>=& \int d^3k'\;(2\pi)^6\delta^3(\textbf{k}_2+\textbf{k}')P(k_2)\delta^3(\textbf{k}_1+\textbf{k}_3-\textbf{k}')P(k_1)\\
    =&(2\pi)^6\delta^3(\textbf{k}_2+\textbf{k}_3+\textbf{k}_1)P_{{}_\zeta}(k_1)P_{{}_\zeta}(k_2).
    \end{align*}
Therefore, adding everything we get
\begin{align}
    \left<\zeta(\textbf{k}_1)\zeta(\textbf{k}_2)\zeta(\textbf{k}_3)\right> &= +\frac{3}{5}f_{{}_{\rm NL}}\frac{1}{(2\pi)^3}\left<\zeta_{{}_{\rm L}}(\textbf{k}_1)\zeta_{{}_{\rm L}}(\textbf{k}_2)\int d^3k'\; \zeta_{{}_{\rm L}}(\textbf{k}_3-\textbf{k}') \; \zeta_{{}_{\rm L}}(\textbf{k}')\right>+{\rm sym.}\nonumber\\
    &=+\frac{6}{5}(2\pi)^3\delta^3(\textbf{k}_1+\textbf{k}_2+\textbf{k}_3)f_{{}_{\rm NL}}\bigg(P_{{}_\zeta}(k_1)P_{{}_\zeta}(k_2) + P_{{}_\zeta}(k_3)P_{{}_\zeta}(k_2) +P_{{}_\zeta}(k_1)P_{{}_\zeta}(k_3) \bigg).
\end{align}
Substituting for dimensionless power spectrum, $\Delta_{{}_\zeta}(k)=P_{{}_\zeta}(k) \frac{k^3}{2\pi^2}$ we can express above expression as:
\begin{align}
     \left<\zeta(\textbf{k}_1)\zeta(\textbf{k}_2)\zeta(\textbf{k}_3)\right>=\frac{3}{10}(2\pi)^7\delta^3(\textbf{k}_1+\textbf{k}_2+\textbf{k}_3)f_{{}_{\rm NL}}\bigg(\frac{\Delta_{{}_\zeta}(k_1)\Delta_{{}_\zeta}(k_2)}{k_1^3k_2^3} + \frac{\Delta_{{}_\zeta}(k_3)\Delta_{{}_\zeta}(k_2)}{k_3^3k_2^3}+ \frac{\Delta_{{}_\zeta}(k_1)\Delta_{{}_\zeta}(k_3)}{k_3^3k_1^3}\bigg)\nonumber
\end{align}
If scale dependence of $\Delta_{{}_\zeta}(k)$ is negligible, $|n_s-1|\ll 1$ then we get: 
\begin{equation}
     \left<\zeta(\textbf{k}_1)\zeta(\textbf{k}_2)\zeta(\textbf{k}_3)\right>=\frac{3}{10}(2\pi)^7\delta^3(\textbf{k}_1+\textbf{k}_2+\textbf{k}_3)f_{{}_{\rm NL}}~[\Delta_{{}_\zeta}(k_1)]^2\bigg(\frac{k_1^3+k_2^3+k_3^3}{k_1^3k_2^3k_3^3}\bigg)\,.\nonumber
\end{equation}
After reordering 
\begin{equation}
     \left<\zeta(\textbf{k}_1)\zeta(\textbf{k}_2)\zeta(\textbf{k}_3)\right>=(2\pi)^7\delta^3(\textbf{k}_1+\textbf{k}_2+\textbf{k}_3)\frac{[\Delta_{{}_\zeta}(k_1)]^2}{k_1^3k_2^3k_3^3}\times \left\{ +\frac{3}{10}f_{{}_{\rm NL}}~\bigg(k_1^3+k_2^3+k_3^3\bigg)\right\}\nonumber
\end{equation}
Therefore, in the case of \eqref{localng} when collecting different contributions to bispectrum as 
\begin{equation}
\left<\zeta(\textbf{k}_1)\zeta(\textbf{k}_2)\zeta(\textbf{k}_3)\right>=(2\pi)^3\delta^3(\textbf{k}_1+\textbf{k}_2+\textbf{k}_3)B(k_1,k_2,k_3)
\end{equation}
we get 
    
    \begin{equation}
    f_{{}_{\rm NL}} = \frac{1}{(2\pi)^4\Delta_{{}_\zeta}^2}B(k_1,k_2,k_3) \left\{\frac{10}{3}\frac{\prod k_i^3}{\sum k_i^3}\right\}. 
\end{equation}
\bibliographystyle{JHEP}
\bibliography{bib2}
\clearpage

\end{document}